\documentclass[journal]{IEEEtran}

\usepackage{cite}
\usepackage{amsmath,amssymb,amsfonts}
\usepackage{bm}
\usepackage{graphicx}
\usepackage{booktabs}
\usepackage{siunitx}
\usepackage{multirow}
\usepackage{array}
\usepackage{url}
\usepackage{algorithm}      %
\usepackage[noend]{algpseudocode}
\usepackage[caption=false,font=footnotesize]{subfig}
\usepackage{flafter}
\usepackage{tikz}
\usetikzlibrary{patterns}
\usepackage[hidelinks]{hyperref}
\usepackage{orcidlink}
\hypersetup{
  pdftitle={A Tutorial on Infrastructure Self-Calibration for Radio-Based Indoor Positioning},
  pdfauthor={Paul Schwarzbach},
  pdfsubject={Preprint},
  pdfkeywords={5G positioning, anchor self-calibration, autocalibration,
    Bluetooth Low Energy (BLE), real-time locating systems (RTLS),
    simultaneous localization and mapping (SLAM), ultra-wideband (UWB),
    WiFi fine timing measurement (FTM)}}

\graphicspath{{figures/}}

\newcommand{\pos}[1]{\mathbf{p}_{#1}}            %
\newcommand{\pose}[1]{\mathbf{x}_{#1}}           %
\newcommand{\meas}[1]{\tilde{d}_{#1}}            %
\newcommand{\edges}{\mathcal{E}}                 %
\newcommand{\edgesAA}{\mathcal{E}_{\mathrm{aa}}} %
\newcommand{\edgesAM}{\mathcal{E}_{\mathrm{am}}} %
\newcommand{\known}{\mathcal{K}}                 %
\newcommand{\knownM}{\mathcal{K}_{\mathrm{m}}}   %
\newcommand{\nodes}{\mathcal{V}}                 %
\newcommand{\nodesA}{\mathcal{V}_{\mathrm{a}}}   %
\newcommand{\nodesM}{\mathcal{V}_{\mathrm{m}}}   %
\newcommand{\graphG}{\mathcal{G}}                %
\newcommand{\allvars}{\mathcal{X}}               %
\newcommand{\rhofun}{\rho}                       %
\newcommand{\facRange}[1]{\psi_{#1}}             %
\newcommand{\facMotion}[1]{\phi_{#1}}            %
\newcommand{\facPrior}[1]{\pi_{#1}}              %
\newcommand{\onelinelonger}{\looseness=1\relax}

\newcommand{\WEfmCauchySpanSetA}{0.63--0.79}
\newcommand{\WEfmGaussSpanSetA}{2.9--3.9}

\newcommand{\WEfmBatchGaussBthree}{0.96}
\newcommand{\WEfmBatchCauchyBthree}{0.22}

\newcommand{\WEfmCauchyFloorSetB}{0.22--0.25}

\newcommand{\WEmhFgCauchyKThreeSetA}{9.66}

\newcommand{\WEmhRobustKThreeSpanSetB}{0.42--0.43}

\newcommand{\WEcrlbMhKthreeBthree}{0.984}
\newcommand{\WEcrlbMhKthreeWeakBthree}{0.941}

\newcommand{\WEslamEpochsK}{3.4}
\newcommand{\WEmhRigidKSetA}{5}
\newcommand{\WEmhRigidKSetB}{4}

\newcommand{\WEctlRandRigidSetB}{16}
\newcommand{\WEctlRandSplitSetB}{11}

\newcommand{\WEctlRandSplitSetA}{9}

\newcommand{\WEctlOracleKThreeSetA}{0.43}

\newcommand{\WEctlSweepQuarterSpanSetA}{0.27--0.32}

\newcommand{\WEoffFmCommonSetB}{0.148}

\newcommand{\WEoffWidth}{0.10}

\newcommand{\WEscaleShareSpanSetB}{70--81}
\newcommand{\WEscaleExpandSpanSetB}{2.5--2.6}

\newcommand{\WEscaleExpandDeviceSetB}{1.1}

\newcommand{\WEresAaMedSetB}{0.27}

\newcommand{\WEnodeSets}{3}

\newcommand{\WEnodeAnchorsSetB}{10}
\newcommand{\WEpfSeeds}{5}

\newcommand{\WEsigMdsAtwentyone}{1.62}

\newcommand{\WEsigMdsBthree}{0.129}

\newcommand{\WEfmEngineSpreadSetB}{0.03}

\newcommand{\WEctlRandCostWrongSetA}{5}

\newcommand{\WEwtAnchors}{10}

\newcommand{\WEwtClassical}{0.26}

\newcommand{\WEwtGaussMaxErr}{2.44}

\newcommand{\WEwtCauchyMaxErr}{0.31}
\newcommand{\WEwtCauchyObsWeight}{0.003}

\newcommand{\WEsynSigmaHw}{0.33}

\newcommand{\WEsynDraws}{20}

\newcommand{\WEsynNlosStartWorse}{2}

\newcommand{\WEfcPairsBthree}{45}

\newcommand{\WEfcCfFullBthree}{30}
\newcommand{\WEfcCfMinBthree}{0.32}

\newcommand{\WEfcCfMaxBthree}{5.92}

\newcommand{\WEfcEkfMinAtwentyone}{0.647}

\newcommand{\WEfcEkfMaxAtwentyone}{1.512}

\newcommand{\WEextSharedSpan}{1.6--2.2}
\newcommand{\WEextSurveyedSpan}{0.69--0.71}

\newcommand{\WEextKernelDiff}{0.02}

\begin{document}

\title{A Tutorial on Infrastructure Self-Calibration\\
for Radio-Based Indoor Positioning}

\author{Paul~Schwarzbach\,\orcidlink{0000-0002-1091-782X}%
\thanks{This work has been submitted to the IEEE for possible publication. Copyright may be transferred without notice, after which this version may no longer be accessible.}%
\thanks{P. Schwarzbach is with the Chair of Transport Systems Information
Technology, TUD Dresden University of Technology, Dresden, Germany
(e-mail: paul.schwarzbach@tu-dresden.de).}}

\markboth{}{Schwarzbach: Infrastructure Self-Calibration for Radio-Based Indoor Positioning}

\maketitle

\begin{abstract}
Radio positioning needs the positions of its anchors, access points, base
stations, or reconfigurable intelligent surfaces, and manual surveying is
costly and error-prone. Calibrating this infrastructure from its own
measurements is one estimation problem, studied in at least four communities
as self-calibration, auto-positioning, anchor self-localization, network
localization, or radio simultaneous localization and mapping. This tutorial
teaches the problem through one factor graph. The measurement topology sets
its factors, the reference information enters as priors, and other
measurement models add nuisance states. The tutorial shows what ranges can
identify, from rigidity to the motion a mobile device must contribute, and
presents multidimensional scaling, nonlinear least squares, recursive
filters, and convex relaxations as inference engines over this graph. One
session of a ten-anchor ultra-wideband deployment is carried through every
step, and recordings from a second, harsher deployment and a public partial
mesh serve as contrasts. The examples illustrate three lessons. On a full
mesh, choose a robust loss and set its scale first, and calibrate with the
engine of the positioning system, since the engines then reach similar
errors. On thinned or partial meshes, test the result at surveyed check
points, because the start decides which minimum a solver reaches. On a
line-of-sight deployment, estimate range offsets, which lower the error floor
that a systematic ranging bias sets and a robust loss cannot remove. A
decision tree and a table of their effects turn them into deployment
guidance. Code and result tables are openly released, and the data are
published separately.
\end{abstract}


\begin{figure}[t]
  \centering
  \includegraphics[width=\columnwidth]{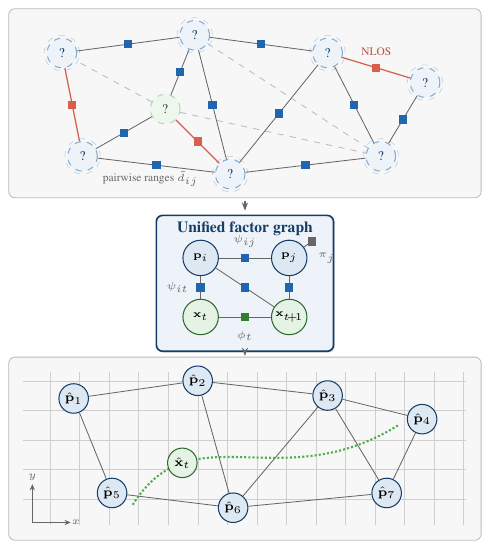}
  \caption{Infrastructure self-calibration from radio ranges among anchors of
    unknown position (marked~?) and a mobile tag (green).
    Circles are variables and squares are range ($\facRange{}$), motion
    ($\facMotion{}$), and prior ($\facPrior{}$) factors. Dashed pairs are out
    of range, red links non-line-of-sight.}
  \label{fig:graphical_abstract}
\end{figure}

\section{Introduction}
\label{sec:introduction}

\IEEEPARstart{L}{ocation} awareness has moved from a convenience to an
operating requirement in settings as different as industrial production,
warehouse logistics, autonomous robots, emergency response, and smart
buildings~\cite{hailu_theories_2024}. Satellite navigation does not reach most
of these settings, because its signals attenuate indoors and degrade in dense or
underground environments. Terrestrial radio systems provide the position fix
instead, built on ultra-wideband (UWB) ranging, WiFi timing measurements,
Bluetooth Low Energy (BLE) beacons, Long Range (LoRa) networks, or cellular
signals~\cite{roy_survey_2021,kosek-szott_indoor_2026,plets_experimental_2018}.
Two decades of work on network and cooperative localization have refined how a
mobile device is positioned from measurements to fixed
nodes~\cite{wymeersch_cooperative_2009,buehrer_collaborative_2018}. Most of
those methods take one assumption as given: the positions of the fixed nodes
are known in advance~\cite{xu_survey_2025}.

Errors in those assumed coordinates do not average out. They propagate into
every position the system reports, with an impact quantified analytically and
experimentally~\cite{qi_calibration_2024}. Angle-based nodes, such as 5G
millimeter-wave base stations or Bluetooth direction-finding locators, also
need a known antenna
orientation~\cite{lu_bayesian_2023,schwarzbach_terminology_2026}. The established way to
obtain the coordinates is geodetic surveying: after installation, each node is
measured with an optical instrument such as a total
station~\cite{hamesse_fast_2024,leica_TS16}. Surveying delivers accuracy, but
it binds trained personnel and equipment to every node, which
makes it the main driver of the deployment cost, as reported for
UWB~\cite{ridolfi_self-calibration_2021}. It is described as time-consuming
and error-prone at the scale of dozens of
anchors~\cite{herbruggen_multihop_2023,weber_automatisierte_2021}, and its result is perishable, since
any relocation or extension of the infrastructure restarts the
procedure~\cite{krapez_anchor_2020}. A side-by-side evaluation of indoor
location systems found their deployment and profiling cost prohibitively
high~\cite{lymberopoulos_realistic_2015}. An industry analysis reports that
most surveyed enterprises had not yet deployed a real-time locating system
and that the cost of deployment and installation can be a bigger barrier than
that of the infrastructure itself, particularly where centimeter-level
accuracy is required~\cite{zignani_5g_2021}.

Infrastructure self-calibration removes the surveyor from this loop. The nodes
range to each other, to a mobile device passing through, or both, and
the deployment estimates its own geometry from those measurements, as
Fig.~\ref{fig:graphical_abstract} sketches. Such proposals reach back
several decades, from array shape calibration with sources at unknown
positions~\cite{rockah_array_1987} and range-only mapping of radio beacons
from a moving robot~\cite{kantor_preliminary_2002} to anchor calibration for
indoor positioning
systems~\cite{zhou_construction_2010,muller_automatic_2011}. UWB networks with
inter-anchor ranging turned the idea into a practical system
component~\cite{hamer_self-calibrating_2018,vashistha_self-calibration_2018},
radio simultaneous localization and mapping (SLAM) lets a mobile agent
map the infrastructure while localizing itself~\cite{gentner_simultaneous_2017},
and the line remains active across
technologies~\cite{corbalan_self-localization_2023,hamesse_fast_2024,tong_mapfi_2023}.

The idea has also reached products. UWB real-time locating kits ship an
auto-positioning routine for up to four anchors in a prescribed
layout~\cite{qorvo_mdek1001,almansa_autocalibration_2020}, and WiFi access
points (APs) locate themselves in the
field~\cite{hpe_self-locating_2025,kosek-szott_indoor_2026}. The shipped
features leave headroom. An independent characterization of one such routine
measured a maximum anchor error of \SI{1.2}{\meter} after a \SI{40}{\second}
procedure, where a re-estimation on identical hardware reached
\SI{0.4}{\meter} in under \SI{3}{\second}~\cite{queralta_uwb_2020}.

The research behind these products is fragmented, starting with its name. The
same operation appears in titles as
self-calibration~\cite{hamer_self-calibrating_2018,vashistha_self-calibration_2018},
anchor self-localization~\cite{corbalan_self-localization_2023},
auto-positioning~\cite{jung_auto-positioning_2022},
autocalibration~\cite{almansa_autocalibration_2020,raspopoulos_3d_2026}, automatic anchor
calibration~\cite{muller_automatic_2011}, or construction of a local anchor
map~\cite{zhou_construction_2010}, so a literature search under any single
label returns a fragment of the relevant
work~\cite{furnas_vocabulary_1987,schwarzbach_terminology_2026}. The scatter
also runs across communities. Signal processing
developed algebraic and closed-form solutions for the self-calibration of
microphone arrays that radio work later
inherited~\cite{crocco_bilinear_2012,kuang_microphone_2013}, and robotics
treats the fixed nodes as landmarks of a SLAM
problem~\cite{kantor_preliminary_2002,gentner_simultaneous_2017}. The
wireless-sensor-network community formulated network localization for large
multi-hop meshes, with a few anchors through distance-vector
propagation~\cite{niculescu_aps_2002} or belief
propagation~\cite{ihler_nonparametric_2005}, and without anchors through
relative maps~\cite{shang_localization_2003,priyantha_anchor-free_2003}.
Navigation and communications estimate transmitter states as part of
cooperative localization and signals-of-opportunity
processing~\cite{wymeersch_cooperative_2009,khalife_opportunistic_2020,kassas_resilient_2022}.
Each community keeps its own notation and canonical baselines, so a newcomer
has to assemble the state of the art across them.

This paper is a tutorial on the geometric calibration of radio
infrastructure, meaning the anchors, APs, base stations, and
reconfigurable intelligent surfaces~\cite{zheng_jrcup_2024} whose positions a
positioning system needs. It aims to let a reader state a deployment as one
factor graph, judge from its edge set what the ranges can identify, choose an
algorithm family and its starting point, and check a calibration before
trusting it. It assumes least-squares estimation and Bayesian filtering at the
level of introductory
texts~\cite{thrun2005probabilistic,brown2012IntroductionRandomSignals}, and
Sections~\ref{sec:formulation} and~\ref{sec:identify} introduce factor
graphs~\cite{dellaert_factor_2017} and graph
rigidity~\cite{aspnes_theory_2006} as far as the tutorial uses them. The
formulation is built on ranges. Time differences, one-way arrival times,
angles, signal strength, and carrier phase enter through their measurement
models (Section~\ref{sec:nuisance}), and the examples of this tutorial use
UWB two-way ranging in two dimensions.

\subsection{Existing Surveys and Tutorials}
\label{sec:related}

The surveys and comparisons closest to this tutorial each center on one strand
or technology. Ridolfi et~al.~\cite{ridolfi_self-calibration_2021} survey UWB
self-calibration together with collaborative localization, and Amjad
et~al.~\cite{amjad_radio_2023} survey radio SLAM. Nurminen
et~al.~\cite{nurminen_survey_2017} survey transmitter localization from signal
strength, and a survey of mobile-anchor-assisted
localization~\cite{han_survey_2016} treats static nodes that a moving node with
known positions localizes. Amaral Simões
et~al.~\cite{amaral_simoes_systematic_2025} compare calibration techniques for
UWB anchors on synthetic data.

Surveys and tutorials of sensor-network, cooperative, and network
localization~\cite{patwari_locating_2005,mao_wireless_2007,wymeersch_cooperative_2009,buehrer_collaborative_2018,win_theoretical_2018}
share the estimation core of the problem but sort the nodes by whether their
positions are known. Wymeersch et~al. distinguish ``agents, which have a
priori unknown states, and anchors, which have known states at all times'',
and either kind may move~\cite{wymeersch_cooperative_2009}. Patwari et~al. call
the sensors with known coordinates reference nodes and list network
calibration among the names of cooperative
localization~\cite{patwari_locating_2005}. Where the unknown nodes are static,
as in sensor networks, the estimation problem is that of the anchor-only
topologies of Section~\ref{sec:taxonomy} with a few positions known. Most of
these works give conditions under which such a network is localizable or
bounds on the error of its
estimate~\cite{patwari_locating_2005,mao_wireless_2007,buehrer_collaborative_2018,win_theoretical_2018},
but none develops the case of a moving device that ranges to static nodes of
unknown position. The acoustic overview of Plinge
et~al.~\cite{plinge_acoustic_2016} combines both kinds of topology with
measured examples, for microphone arrays rather than radio infrastructure.

Surveys and tutorials of indoor, cellular, and WiFi
positioning~\cite{laoudias_survey_2018,zafari_survey_2019,italiano_tutorial_2025,kosek-szott_indoor_2026}
describe primitives and standards and treat the calibration of the
infrastructure at most as one topic among many. A companion
review~\cite{schwarzbach_terminology_2026} examined how the literature names
the calibration problem, showed that the labels largely describe one
estimation task, and introduced the framework of measurement topologies and
reference information that Section~\ref{sec:taxonomy} restates. None of these
works teaches the calibration of radio infrastructure across its topologies,
from identifiability to the choice of an algorithm, with examples on measured
data and released code.

\subsection{Contributions and Organization}

This tutorial makes three contributions.
\begin{itemize}
  \item \textbf{One formulation across communities.} Range-based calibration
    is stated as one factor graph, whose factor structure follows the
    measurement topology and whose priors carry the reference information,
    while other primitives add nuisance states. Rigidity and its test,
    the motion conditions of the mobile topologies, and the Gaussian bound
    state what the ranges can identify. The algorithm families appear as
    approximate inference engines over this graph, each with its own data
    interface and its need for a starting point
    (Sections~\ref{sec:formulation} to~\ref{sec:algorithms}).
  \item \textbf{A path from deployment to method.} A decision tree leads from
    what a practitioner knows to a topology, an engine, and the checks a
    calibration needs, and each leaf names the examples it rests on
    (Section~\ref{sec:decision_tree}).
  \item \textbf{Reference implementations and examples on open measured
    data.} One session of a
    \WEwtAnchors-anchor UWB deployment is carried through every step, and a
    second deployment rich in non-line-of-sight (NLOS) propagation and a
    public partial mesh serve as contrasts. Section~\ref{sec:worked_examples}
    draws three lessons for practice from them, each with its effect on the
    anchor error. The data of both deployments are published as a separate
    dataset~\cite{schwarzbach_uwbasc_descriptor_2026,schwarzbach_uwbasc_2026}.
    The released code
    recomputes the result tables of every example~\mbox{(Section~\ref{sec:records})}.
\end{itemize}
The formulation builds on factor graphs for robot
perception~\cite{dellaert_factor_2017} and on their use in cooperative
localization~\cite{wymeersch_cooperative_2009} and multipath
SLAM~\cite{leitinger_scalable_2018}. The framework of topologies and reference
information, and the distinction between self-calibration and radio SLAM by
the treatment of the trajectory, come from the companion
review~\cite{schwarzbach_terminology_2026}. No new estimation algorithm is
claimed, and the examples demonstrate known principles on measured data.

The rest of the paper is structured as follows. Section~\ref{sec:background}
states the problem and introduces the running example.
Section~\ref{sec:formulation} formulates the problem as one factor graph,
Section~\ref{sec:identify} shows what its ranges can identify, and
Section~\ref{sec:algorithms} presents the inference engines.
Section~\ref{sec:worked_examples} runs them on measured data, and
Section~\ref{sec:decision_tree} turns the results into a decision tree.
Section~\ref{sec:discussion} discusses open
problems, and
Section~\ref{sec:conclusion} concludes. Readers who need a method for a
concrete deployment can start from the decision tree and follow its pointers
back.

\section{The Problem and the Running Example}
\label{sec:background}

\subsection{What Is Calibrated}

Radio-based positioning rests on a division of roles between fixed and mobile
nodes. A set of static nodes, the infrastructure, occupies known locations and
provides the spatial reference, and one or more mobile devices infer their own
positions from measurements to that infrastructure. In this tutorial,
infrastructure means the anchors of dedicated positioning systems and the
APs, base stations, and reconfigurable intelligent surfaces of
communication networks. The names differ by
community, so that anchors in UWB, satellites in the \mbox{omlox} standard
for real-time locating systems~\cite{pno_omlox_2024}, APs in WiFi, beacons in BLE, base
stations (gNodeBs) in 5G, and gateways in LoRa networks all fill the same
reference role, while tags, agents, and user equipment name the mobile side.
Here every infrastructure node is called an anchor.

Infrastructure calibration replaces the manual survey. Given measurements among the
nodes and, optionally, a small set of surveyed positions, it estimates the
coordinates of the infrastructure in a common frame. The operation is the dual of
node localization. Localization holds the infrastructure fixed and solves for a
mobile position, whereas calibration treats the infrastructure positions as the
unknowns and recovers them from the
measurements~\cite{schwarzbach_terminology_2026}. Both read the same
measurements through the same geometry, so one estimator template covers them
once the roles of known and unknown are assigned.
Positioning a mobile device against known infrastructure is the mode that
calibration supports, not an instance of it. Nuisance states such as device
delays, clock offsets, or anchor orientations belong to the problem when they
are estimated jointly with the
geometry~\cite{hol_ultra-wideband_2010,preter_range_2019,lu_bayesian_2023}.
Neighboring strands contribute methods that transfer. Acoustic and ultrasound
self-calibration shares the estimation core but not the channel
physics~\cite{runge_precise_2011}, and cooperative localization, delay
calibration of infrastructure with known
positions~\cite{horvath_calibration_2017}, and signals of opportunity share
parts of the graph.

\subsection{Measurement Topology and Reference Information}
\label{sec:taxonomy}\label{sec:scoping_review}

The framework of this subsection comes from the companion
review~\cite{schwarzbach_terminology_2026} and is restated with the mobile
topologies defined by their edge sets and with a mobile-referenced regime. Its
organizing question is structural: how are the measurement constraints
arranged, before any algorithm is chosen? The answer is the measurement
topology, the edge set $\edges$ of the measurement graph
$\graphG = (\nodes, \edges)$, whose vertices $\nodes = \nodesA \cup \nodesM$
divide into infrastructure nodes $\nodesA$ and mobile nodes $\nodesM$.
Throughout, topology names this measurement edge set, not the communication
links to which the sensor-network literature attaches the same word. The
companion review found four topologies recurring across the literature under
otherwise unrelated vocabularies. This tutorial renames its SLAM-based
topology mobile-only, since here the edge set alone defines a topology and the
treatment of the trajectory is a separate choice (Section~\ref{sec:spectrum}).

\textbf{Fully meshed.} Every infrastructure pair exchanges a direct measurement,
producing a complete anchor--anchor edge set
$\edgesAA = \{(i,j) : i,j \in \nodesA,\, i < j\}$ with no mobile node,
$\nodesM = \emptyset$. Among the anchor-only topologies this configuration
yields the strongest constraints and needs no moving participant. It requires
$\binom{N}{2}$ pairwise measurements among $N$ anchors and communication
between all anchor pairs, which complicates the transmission scheme and raises
the power consumption~\cite{shi_anchor_2019}. The full mesh is the setting
of anchor self-calibration systems that range
peer-to-peer~\cite{hamer_self-calibrating_2018,krapez_anchor_2020,jung_auto-positioning_2022}
and of early anchor-map constructions~\cite{zhou_construction_2010}.

\textbf{Multi-hop.} When not all infrastructure pairs lie within range, the
anchor--anchor set is incomplete,
$\edgesAA \subset \{(i,j) : i,j \in \nodesA,\, i < j\}$, and calibration must
propagate through chains of intermediate nodes, with error accumulating along
each
hop~\cite{herbruggen_multihop_2023,muller_automatic_2011,niculescu_aps_2002,shang_localization_2003}.
Multi-hop and fully meshed share the anchor-only character and differ only in
edge density, a difference that can cost the graph its rigidity
(Section~\ref{sec:rigidity}).

\textbf{Mobile-assisted.} A device ranges to the infrastructure on top of the
inter-anchor links, whether it moves along a path or stands at a few points,
so the edge set is mixed,
$\edges \subseteq (\nodesA \times \nodesM) \cup
\edgesAA$ with
$\edgesAA \neq \emptyset$~\cite{yudanto_device-level_2023,preter_range_2019}.
The inter-anchor mesh need not be complete.

\textbf{Mobile-only.} A moving device ranges to the infrastructure and the
anchors do not range among themselves, leaving a purely bipartite edge set
$\edges \subseteq \nodesA \times \nodesM$. Two bodies of work share this
edge set. Radio SLAM binds the device positions with a motion model
and recovers the trajectory together with a map of landmarks, the framing
inherited from robotics~\cite{kantor_preliminary_2002,menegatti_range-only_2009,gentner_simultaneous_2017}.
Bipartite self-calibration from time-of-arrival (ToA) or two-way ranges needs
no motion model
and treats the device positions as nuisance
unknowns~\cite{zhao_multiepoch_2025}. Batstone
et~al.~\cite{batstone_robust_2016} start without one and add a smoothness
prior that penalizes the acceleration of the trajectory only in their final
refinement. Both belong to the
same topology, because only the absent inter-anchor links separate it from the
mobile-assisted case.

The four definitions take a single mobile device, whose poses at the
epochs form $\nodesM$. Section~\ref{sec:spectrum} treats the
mobile--mobile edges of cooperative settings as an extension.

The second dimension, independent of topology, records which variables carry
prior position knowledge. Its regimes adapt the
blind, semi-blind, and non-blind terminology of sensor-data
calibration~\cite{barcelo-ordinas_self-calibration_2019} to reference
positions. Let
$\known \subseteq \nodesA$ collect the infrastructure nodes whose positions are
given, exactly or approximately, and $\knownM$ the epochs at which an external reference
gives the mobile pose. Blind calibration takes
$\known = \knownM = \emptyset$ and recovers the geometry only
up to a rigid transformation, since range measurements are invariant under
translation, rotation, and reflection of the whole
configuration~\cite{aspnes_theory_2006,ash_relative_2008}. Semi-blind
calibration fixes a subset,
$1 \le |\known| < N$~\cite{corbalan_self-localization_2023}, and non-blind
calibration fixes
all nodes, $|\known| = N$, which leaves clocks, antenna delays, and ranging
biases to calibrate. Pose priors fix the
frame in the same way when an external reference supplies the trajectory,
for example a receiver aided by a global navigation satellite system
(GNSS)~\cite{kassas_resilient_2022} or a
laser-scanning inertial rig~\cite{hamesse_fast_2024}. With no anchor pinned,
$\known = \emptyset$ and $\knownM \neq \emptyset$, the regime is
mobile-referenced. The anchor side names the regime when both kinds of prior
are present. Matching the recovered geometry to a digital map after
estimation~\cite{weber_automatisierte_2021} is an alignment of a blind
estimate and not a prior.

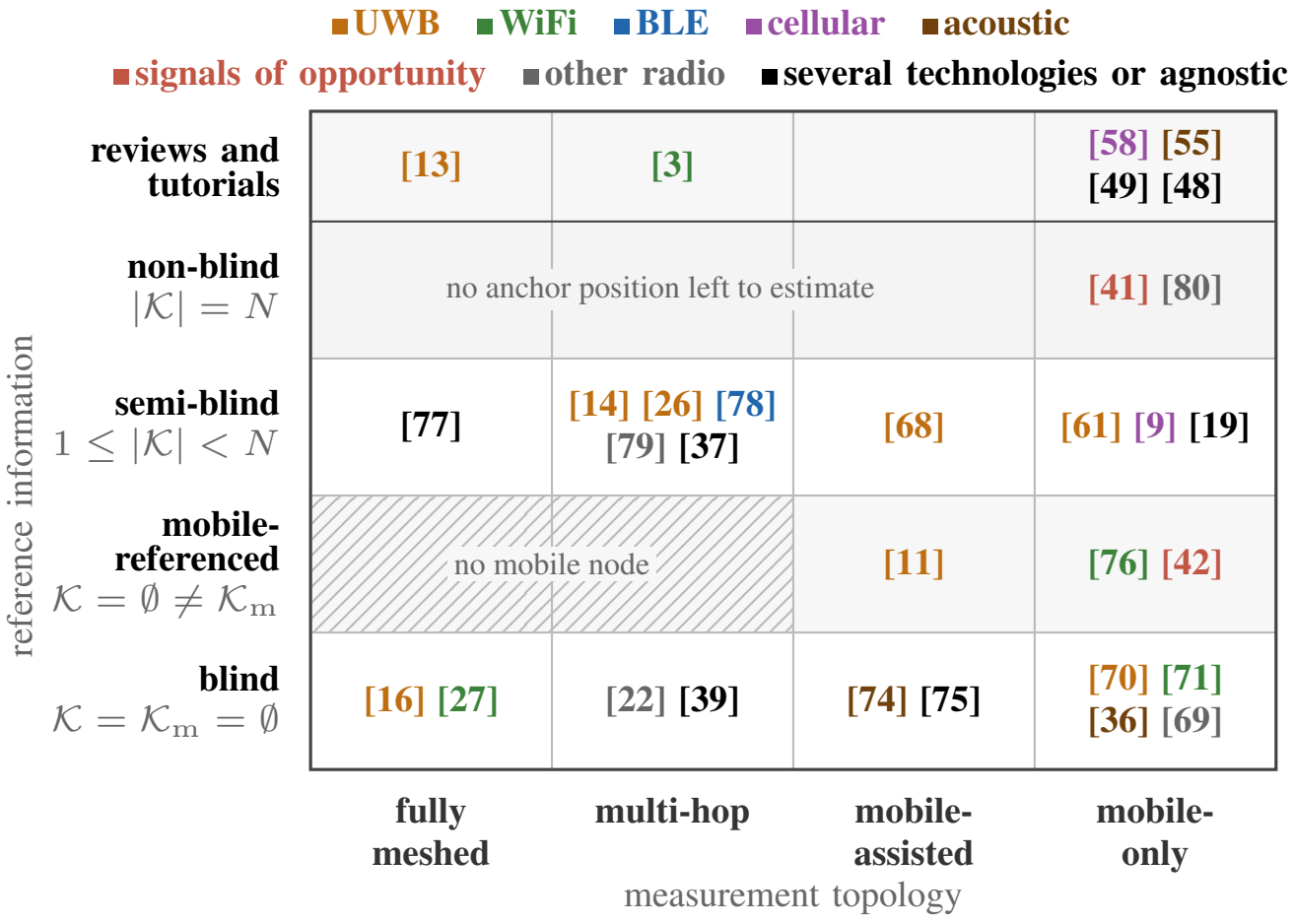
\begin{figure}[!htbp]
  \centering
  \definecolor{tsBlue}{HTML}{2166AC}
\definecolor{tsLightBlue}{HTML}{67A9CF}
\definecolor{tsRed}{HTML}{D6604D}
\definecolor{tsGreen}{HTML}{4DAF4A}
\definecolor{tsPurple}{HTML}{984EA3}
\definecolor{tsOrange}{HTML}{E08214}
\definecolor{tsTeal}{HTML}{1A8A6A}
\definecolor{tsBgLight}{HTML}{F7F7F7}
\definecolor{tsBgMedium}{HTML}{E8E8E8}
\definecolor{tsTxtDark}{HTML}{333333}
\definecolor{tsTxtMedium}{HTML}{666666}
\definecolor{tsBorderDark}{HTML}{444444}
\definecolor{tsBorderLight}{HTML}{BBBBBB}
\colorlet{techUWB}{tsOrange!85!black}
\colorlet{techWiFi}{tsGreen!75!black}
\colorlet{techBLE}{tsBlue}
\colorlet{techCell}{tsPurple}
\colorlet{techSoO}{tsRed!90!black}
\colorlet{techAcoustic}{tsOrange!50!black}%
\colorlet{techOther}{tsTxtMedium}
\colorlet{techAgnostic}{black}
\newcommand*{\tswork}[2]{\textcolor{#1}{\bfseries\citeleft\citen{#2}\citeright}}
\newcommand*{\tskey}[2]{\textcolor{#1}{\rule{0.42em}{0.42em}\,\bfseries #2}}
\begin{tikzpicture}[x=1cm, y=1cm,
    font=\fontsize{7}{8.2}\selectfont,
    cell/.style={align=center, text width=1.50cm, inner sep=0pt},
    note/.style={font=\fontsize{6}{7}\selectfont, text=tsTxtMedium,
      inner sep=1.2pt}]
  \def\colw{1.62}
  \def\bandh{0.92}
  \def\revh{0.74}
  \pgfmathsetmacro{\yreg}{4*\bandh}
  \pgfmathsetmacro{\ytop}{4*\bandh+\revh}
  \pgfmathsetmacro{\xend}{4*\colw}

  \fill[tsBgLight] (0, \bandh) rectangle (\xend, 2*\bandh);
  \fill[tsBgMedium!50] (0, 3*\bandh) rectangle (\xend, \yreg);
  \fill[tsBgLight] (0, \yreg) rectangle (\xend, \ytop);
  \fill[pattern=north east lines, pattern color=tsBorderLight]
    (0, \bandh) rectangle (2*\colw, 2*\bandh);

  \foreach \i in {1,2,3}
    \draw[tsBorderLight, line width=0.4pt] (\i*\colw, 0) -- (\i*\colw, \ytop);
  \foreach \j in {1,2,3}
    \draw[tsBorderLight, line width=0.4pt] (0, \j*\bandh) -- (\xend, \j*\bandh);
  \draw[tsBorderDark, line width=0.4pt] (0, \yreg) -- (\xend, \yreg);
  \draw[tsBorderDark, line width=0.6pt] (0, 0) rectangle (\xend, \ytop);

  \begin{scope}[every node/.style={anchor=north, align=center,
      font=\fontsize{7}{8.2}\selectfont\bfseries, text=tsTxtDark}]
    \node at (0.5*\colw, -0.06) {fully\\meshed};
    \node at (1.5*\colw, -0.06) {multi-hop};
    \node at (2.5*\colw, -0.06) {mobile-\\assisted};
    \node at (3.5*\colw, -0.06) {mobile-\\only};
  \end{scope}
  \node[text=tsTxtMedium] at (2*\colw, -0.86) {measurement topology};

  \begin{scope}[every node/.style={anchor=east, align=right}]
    \node at (-0.08, 0.5*\bandh) {\textbf{blind}\\[-0.5pt]
      {\color{tsTxtMedium}$\mathcal{K}=\mathcal{K}_{\mathrm{m}}=\emptyset$}};
    \node at (-0.08, 1.5*\bandh) {\textbf{mobile-}\\[-1.5pt]\textbf{referenced}\\[-0.5pt]
      {\color{tsTxtMedium}$\mathcal{K}=\emptyset\neq\mathcal{K}_{\mathrm{m}}$}};
    \node at (-0.08, 2.5*\bandh) {\textbf{semi-blind}\\[-0.5pt]
      {\color{tsTxtMedium}$1\le|\mathcal{K}|<N$}};
    \node at (-0.08, 3.5*\bandh) {\textbf{non-blind}\\[-0.5pt]
      {\color{tsTxtMedium}$|\mathcal{K}|=N$}};
    \node at (-0.08, \yreg+0.5*\revh) {\textbf{reviews and}\\[-1.5pt]\textbf{tutorials}};
  \end{scope}
  \node[text=tsTxtMedium, rotate=90, anchor=south, inner sep=0pt]
    at (-1.86, 2*\bandh) {reference information};

  \node[anchor=south, align=center, inner sep=0pt] at (0.5*\xend-0.62, \ytop+0.12)
    {\tskey{techUWB}{UWB}\quad \tskey{techWiFi}{WiFi}\quad
     \tskey{techBLE}{BLE}\quad \tskey{techCell}{cellular}\quad
     \tskey{techAcoustic}{acoustic}\\[1.5pt]
     \tskey{techSoO}{signals of opportunity}\quad
     \tskey{techOther}{other radio}\quad
     \tskey{techAgnostic}{several technologies or agnostic}};

  \node[note, fill=tsBgLight] at (1*\colw, 1.5*\bandh) {no mobile node};
  \node[note, fill=tsBgMedium!50] at (1.45*\colw, 3.5*\bandh)
    {no anchor position left to estimate};

  \node[cell] at (0.5*\colw, 0.5*\bandh)
    {\tswork{techUWB}{krapez_anchor_2020} \tswork{techWiFi}{tong_mapfi_2023}};
  \node[cell] at (1.5*\colw, 0.5*\bandh)
    {\tswork{techOther}{muller_automatic_2011} \tswork{techAgnostic}{shang_localization_2003}};
  \node[cell] at (2.5*\colw, 0.5*\bandh)
    {\tswork{techAcoustic}{djugash_range-only_2006} \tswork{techAgnostic}{mekonnen_self-calibration_2015}};
  \node[cell] at (3.5*\colw, 0.5*\bandh)
    {\tswork{techUWB}{zhao_multiepoch_2025} \tswork{techWiFi}{batstone_robust_2016}
     \tswork{techAcoustic}{kuang_microphone_2013} \tswork{techOther}{menegatti_range-only_2009}};
  \node[cell] at (2.5*\colw, 1.5*\bandh)
    {\tswork{techUWB}{hamesse_fast_2024}};
  \node[cell] at (3.5*\colw, 1.5*\bandh)
    {\tswork{techWiFi}{raja_wifi-rtt_2024} \tswork{techSoO}{kassas_resilient_2022}};
  \node[cell] at (0.5*\colw, 2.5*\bandh)
    {\tswork{techAgnostic}{patwari_relative_2003}};
  \node[cell] at (1.5*\colw, 2.5*\bandh)
    {\tswork{techUWB}{herbruggen_multihop_2023} \tswork{techUWB}{corbalan_self-localization_2023}
     \tswork{techBLE}{li_rss-based_2026} \tswork{techOther}{costa_distributed_2006}
     \tswork{techAgnostic}{niculescu_aps_2002}};
  \node[cell] at (2.5*\colw, 2.5*\bandh)
    {\tswork{techUWB}{yudanto_device-level_2023}};
  \node[cell] at (3.5*\colw, 2.5*\bandh)
    {\tswork{techUWB}{leitinger_scalable_2018} \tswork{techCell}{lu_bayesian_2023}
     \tswork{techAgnostic}{rockah_array_1987}};
  \node[cell] at (3.5*\colw, 3.5*\bandh)
    {\tswork{techSoO}{khalife_opportunistic_2020} \tswork{techOther}{pohlmann_localization_2022}};
  \node[cell] at (0.5*\colw, \yreg+0.5*\revh)
    {\tswork{techUWB}{ridolfi_self-calibration_2021}};
  \node[cell] at (1.5*\colw, \yreg+0.5*\revh)
    {\tswork{techWiFi}{kosek-szott_indoor_2026}};
  \node[cell] at (3.5*\colw, \yreg+0.5*\revh)
    {\tswork{techCell}{italiano_tutorial_2025} \tswork{techAcoustic}{plinge_acoustic_2016}
     \tswork{techAgnostic}{nurminen_survey_2017} \tswork{techAgnostic}{amjad_radio_2023}};
\end{tikzpicture}
  \caption{The two-dimensional classification space with works for further
    study. Hatched cells cannot occur.}
  \label{fig:two_axis_space}
\end{figure}

Measurement topology and reference information span the two-dimensional
classification of Fig.~\ref{fig:two_axis_space}. The topology fixes which
measurement edges are present, the reference information fixes which
variables carry priors, and the two combine freely, except that the
mobile-referenced regime presupposes a mobile node. A fully meshed system may
be blind or seeded, and a mobile-only system may run without priors, with a
few known landmarks, or against a referenced trajectory. Methods with the
same two coordinates solve the same structural problem, whatever their names,
so the coordinates tell what a method can and cannot recover.
Fig.~\ref{fig:two_axis_space} places up to two works for further study per
technology and topology in the cell of their regime, with the reviews and
tutorials above the grid. They are drawn from a literature search across the
four communities and from the references of this tutorial.

\subsection{Technologies, Primitives, and Node Roles}

The radio technology a system uses is the most visible way to classify it. The
companion review found the calibration literature heavily UWB, with BLE and 5G
absent from its included papers~\cite{schwarzbach_terminology_2026}. Technology
alone does not decide how a system calibrates, which depends on two
attributes. The first is the measurement primitive the radio exposes, which
may be two-way ranging, one-way ToA or its time difference (TDoA), received
signal strength (RSS), angle of
arrival (AoA) or of departure (AoD), or carrier phase. Arrival times, angles,
and multipath components are extracted from the channel state information
(CSI) of WiFi and cellular links or the channel impulse
response (CIR) of UWB, on which multipath-based radio
SLAM~\cite{leitinger_scalable_2018} and channel
charting~\cite{studer_channel_2018} also build
(Section~\ref{sec:algorithms}). The second is
node-role symmetry: whether any node can measure to any other, termed
peer-to-peer here, or whether measurements run only between the infrastructure
and its clients, termed infrastructure-to-client. Technology is a proxy for
these attributes because a radio usually ships one dominant primitive and one
deployment pattern.

In two-way ranging, the responder answers a message of the initiator after a
reply time, and the round trip less the reply time is twice the time of
flight, so the two devices need no common
clock~\cite{ieee_802_15_4_2024,dardari_ranging_2009}. Their clocks run at
slightly different rates, however, and in this single-sided exchange the
range error grows with the rate difference times the long reply
time~\cite{dardari_ranging_2009}. A third message makes the exchange
double-sided and removes the reply time from the
error~\cite{yudanto_device-level_2023}, in an asymmetric form in Piavanini
et~al.~\cite{piavanini_calibration_2022}. Both works also calibrate the
antenna delays of the transmit and receive chains. These add a constant per
device to every round trip, which Section~\ref{sec:nuisance} models as an
offset state.

Standards fix which primitives and roles a technology exposes, and
Table~\ref{tab:tech_primitives} summarizes them. IEEE
802.15.4~\cite{ieee_802_15_4_2024} defines UWB ranging between any two
devices, including the secured and one-way modes introduced by
802.15.4z~\cite{ieee_802_15_4z_2020}, so an
anchor set can measure itself. WiFi fine
timing measurement (FTM) of IEEE 802.11~\cite{ieee_802_11_2024} is two-way
ranging as well and supports
the same arrangement in principle~\cite{banin_scalable_2019}. Ranging between
APs depends on vendor support, however, and research
calibration has mostly relied on client-assisted
geometry~\cite{batstone_robust_2016,raja_wifi-rtt_2024}. Commercial
self-locating APs range among themselves and take GNSS fixes at units on the
building perimeter as
seeds~\cite{kosek-szott_indoor_2026,hpe_self-locating_2025}, and research
systems recover AP geometry from CSI exchanged
among the APs~\cite{tong_mapfi_2023}. LoRa networks locate end
devices from the RSS of their uplinks or from TDoA at
gateways synchronized by GNSS~\cite{plets_experimental_2018}, and a LoRa
transceiver for the \SI{2.4}{\giga\hertz} band adds two-way time-of-flight
ranging~\cite{muller_outdoor_2021}. The 5G New Radio (NR) positioning of
the Third Generation Partnership Project (3GPP) combines one-way time
differences and angles with two-way multi-round-trip time (multi-RTT), and
base stations do not range to each other in these
modes~\cite{3gpp_ts_38305_2026,italiano_tutorial_2025}. Where the parameters
of the transmission-reception points (TRPs) of a network are unknown or wrong,
antenna orientation has to be recovered alongside
position~\cite{lu_bayesian_2023}, which favors joint estimation of
infrastructure and a moving user.

\begin{table*}[!t]
  \caption{Radio technologies as proxies for measurement primitives and node
    roles.}
  \label{tab:tech_primitives}
  \centering
  \small
  \begin{tabular}{@{}>{\raggedright}p{2.0cm}>{\raggedright}p{2.8cm}%
    >{\raggedright}p{3.4cm}>{\raggedright}p{3.5cm}%
    >{\raggedright\arraybackslash}p{4.4cm}@{}}
    \toprule
    Technology & Standard & Primitives & Exposed node roles & Topology tendency \\
    \midrule
    UWB &
      IEEE 802.15.4-2024~\cite{ieee_802_15_4_2024} &
      Two-way ranging, TDoA, AoA, RSS, CIR &
      Peer-to-peer: any node ranges to any other &
      All four \\
    WiFi &
      IEEE 802.11-2024~\cite{ieee_802_11_2024} &
      Two-way ranging (FTM), RSS, CSI, AoA &
      Client-to-AP; inter-AP ranging and CSI vendor-specific &
      Mobile-only; anchor-only or mobile-assisted where APs
      measure each other \\
    BLE &
      Bluetooth Core 6.3~\cite{bluetooth_core_2026} &
      RSS, AoA/AoD (direction finding), Channel Sounding &
      Peer-to-peer RSS and ranging; AoA/AoD from locators &
      RSS meshes possible~\cite{li_rss-based_2026}; ranging meshes possible \\
    Cellular (5G~NR) &
      3GPP TS 38.305, Rel-19~\cite{3gpp_ts_38305_2026} &
      TDoA, AoA/AoD, multi-RTT, carrier phase, CSI &
      Infrastructure-to-client; sidelink between user devices &
      Mobile-only: joint user and infrastructure estimation \\
    Signals of opportunity &
      none &
      Pseudorange, Doppler, carrier phase &
      Roles reversed: unknown transmitters, moving reference receiver &
      Mobile-only; semi-blind, or non-blind when only clocks are estimated \\
    \bottomrule
  \end{tabular}
\end{table*}

Which of the four topologies a system can form is therefore decided by the
measurement primitive and the node roles, not by the technology label.
Peer-to-peer two-way ranging admits the anchor-only topologies, fully meshed
when connectivity is complete and multi-hop when it is not, because the
infrastructure can measure itself. A client-facing primitive, such as one-way
TDoA to passive tags or client-initiated two-way ranging, supplies no
inter-anchor edges unless the anchors also measure each other, as TDoA
networks do for synchronization or
self-calibration~\cite{zhou_construction_2010,wang_automatic_2020}. Such a
system is mobile-only, or mobile-assisted where those inter-anchor
measurements exist, and a moving node carries the
constraints~\cite{hol_ultra-wideband_2010}. This is why UWB tends to appear in
meshed calibration~\cite{hamer_self-calibrating_2018,corbalan_self-localization_2023}
and 5G in SLAM-like formulations~\cite{lu_bayesian_2023}. BLE beacons broadcast
advertisements, so their positions have been estimated from the RSS that a
moving device records~\cite{chen_locating_2017,nurminen_survey_2017}.
Channel Sounding instead ranges two-way between two connected devices that
both support it, either of which may initiate~\cite{bluetooth_core_2026}, and
an evaluation on commercial hardware judged it suited to keyless
entry~\cite{wieme_performance_2025}. Two anchors that support Channel Sounding
can therefore range to each other, so BLE anchors can form the anchor-only
topologies from ranges, not only from RSS.
The decision tree of
Section~\ref{sec:decision_tree} therefore enters through the primitive and
asks first whether the anchors can measure each other.

\subsection{The Running Example}
\label{sec:running_example}\label{sec:datasets}

The examples use the line-of-sight (LOS) deployment of the UWB Anchor
Self-Calibration Dataset
(UWB-ASC)~\cite{schwarzbach_uwbasc_descriptor_2026,schwarzbach_uwbasc_2026},
in which \WEnodeAnchorsSetB{} anchors in an industrial hall, surveyed by a
total station, perform two-way ranging with DW1000 modules. Its static session
\texttt{s03} is the running example, and in its walks \texttt{s01} and
\texttt{s02} the motion-capture truth of the moving tag serves the evaluation
only. The NLOS deployment, among obstructions in the same hall, is the
contrast. In its quasi-static sessions \texttt{s01} to \texttt{s20} a tag
stands on a different surveyed marker in each, and in its static sessions
\texttt{s21} and \texttt{s22} all nodes stand at surveyed positions. Before
UWB-ASC~\cite{schwarzbach_uwbasc_descriptor_2026}, only these static sessions
had been analyzed, in the author's
study~\cite{jung_uncertainty_propagation_2025}. A third layout comes from the
public TWR-CLOUD-Bern dataset of another
group~\cite{xenakis_cooperative_2026,xenakis_twr-cloud-bern_2026}, in which
static nodes of a later chip generation range two-way with each other, at
positions from a laser scan. Its evaluation scan, which the dataset holds out
for assessment, forms a globally rigid partial mesh
(Section~\ref{sec:rigidity}). All layouts are near-planar, so the tutorial
works in two dimensions.

\textit{Code and data.}
\phantomsection\label{sec:records}
The examples use the measured sessions of both deployments and the public
scan, which a reader of UWB-ASC converts, and further public range recordings
join through a reader of their own. The range simulator of UWB-ASC draws
synthetic ranges on its surveyed layouts with error components that a user
sets~\cite{schwarzbach_uwbasc_descriptor_2026}, so the examples also run on
controlled errors. Every session, measured, converted, or synthetic, is one
measurement graph whose nodes carry their roles and positional priors and
whose edges are the directed ranges, with the truth of a tag where one
exists~\cite{schwarzbach_uwbasc_descriptor_2026}, and masks select the views
of Section~\ref{sec:masking}. The release implements the engines that the
algorithm boxes of Section~\ref{sec:algorithms} state. Every implementation
reads a view through one call, \texttt{run}, and returns anchor estimates in
the frame it converged in, so a new method joins the toolkit by implementing
this one signature. The evaluation scores each estimate as
Section~\ref{sec:worked_examples} describes, three leaves of the decision tree
in Section~\ref{sec:decision_tree} name release methods, and the release holds
this code and the result tables, one row per estimate (see Data and Code
Availability).

\textit{Pitfalls.}
\begin{itemize}
  \item A method's name says little about what it can recover, since the
    communities name one problem differently. Its topology and reference
    information place it (Fig.~\ref{fig:two_axis_space}).
  \item Not every radio can measure between its anchors. The primitive and
    node roles decide which topologies a deployment can form and which
    methods apply.
\end{itemize}

\section{One Factor Graph}
\label{sec:formulation}

Table~\ref{tab:notation} collects the main symbols of
Sections~\ref{sec:formulation} to~\ref{sec:algorithms}.

\begin{table}[!htbp]
  \caption{Main symbols of Sections~\ref{sec:formulation}
    to~\ref{sec:algorithms}.}
  \label{tab:notation}
  \centering
  \footnotesize
  \setlength{\tabcolsep}{4pt}
  \begin{tabular}{@{}>{\raggedright}p{1.75cm}>{\raggedright}p{4.6cm}%
    >{\raggedright\arraybackslash}p{1.9cm}@{}}
    \toprule
    Symbol & Meaning & Defined in \\
    \midrule
    $d$ & spatial dimension, 2 or 3 & \eqref{eq:measurement_model} \\
    $\graphG = (\nodes, \edges)$ & measurement graph, nodes and edges &
      Sec.~\ref{sec:taxonomy} \\
    $\nodesA$, $\nodesM$ & anchor nodes, mobile nodes (poses) &
      Sec.~\ref{sec:taxonomy} \\
    $N$ & number of anchors, $|\nodesA|$, or of the nodes of a view in
      Algorithms~\ref{alg:mds} to~\ref{alg:grid} &
      Sec.~\ref{sec:taxonomy}, Algorithm~\ref{alg:mds} \\
    $\edgesAA$, $\edgesAM$ & anchor--anchor, anchor--mobile edges &
      Sec.~\ref{sec:taxonomy}, \eqref{eq:factorization} \\
    $\known$, $\knownM$ & seed anchors, referenced epochs &
      Sec.~\ref{sec:taxonomy} \\
    $\pos{i}$ & position of node~$i$ & \eqref{eq:measurement_model} \\
    $\pose{t}$, $T$ & pose at epoch~$t$, number of epochs &
      Sec.~\ref{sec:factor_graph} \\
    $\allvars$ & unknown anchor positions and poses &
      Sec.~\ref{sec:factor_graph} \\
    $\meas{ij}$ & range measured between $i$ and~$j$ &
      \eqref{eq:measurement_model} \\
    $\epsilon_{ij}$ & range error & \eqref{eq:measurement_model} \\
    $r_{ij}$ & range residual & \eqref{eq:map}, Table~\ref{tab:kernels} \\
    $u_{ij}$ & scaled residual $r_{ij}/\sigma$ &
      Sec.~\ref{sec:robust_loss}, Table~\ref{tab:kernels} \\
    $\sigma$ & noise scale of a session & \eqref{eq:range_factor} \\
    $\rhofun$, $w$ & kernel (loss), its weight &
      \eqref{eq:robust_argmin}, Table~\ref{tab:kernels} \\
    $\facRange{ij}$, $\facMotion{t}$ & range factor, motion factor &
      \eqref{eq:factorization}, \eqref{eq:range_factor} \\
    $\facPrior{k}$, $\boldsymbol{\Sigma}_k$ & unary prior, its covariance &
      Sec.~\ref{sec:factor_graph} \\
    $b$, $\delta_i$ & common and per-device range offsets &
      Table~\ref{tab:nuisance_states} \\
    $\mathbf{F}$, $\mathbf{J}$ & Fisher information, range Jacobian &
      \eqref{eq:fim} \\
    $\kappa$ & neighbor count of a thinned view & Sec.~\ref{sec:masking} \\
    \bottomrule
  \end{tabular}
\end{table}

\subsection{The Pairwise Estimation Problem}

Let
$\pos{i} \in \mathbb{R}^d$, $d \in \{2,3\}$, denote the position of node~$i$,
let $\meas{ij}$ denote a noisy range measurement between nodes $i$ and~$j$,
and let $\edges$ collect the node pairs for which measurements exist. The
measurement model~\cite{schwarzbach_terminology_2026} is
\begin{equation}
\label{eq:measurement_model}
  \meas{ij} = \|\pos{i} - \pos{j}\| + \epsilon_{ij},
  \quad (i,j) \in \edges ,
\end{equation}
with range error $\epsilon_{ij}$, and the positions follow from minimizing
the residuals under a kernel $\rhofun$, also called a loss,
\begin{equation}
\label{eq:robust_argmin}
  \hat{\mathbf{p}}_1, \ldots, \hat{\mathbf{p}}_N
  = \arg\min_{\pos{1},\ldots,\pos{N}}
    \sum_{(i,j) \in \edges}
    \rhofun\!\left(\meas{ij} - \|\pos{i} - \pos{j}\|\right),
\end{equation}
subject to $\pos{k} = \pos{k}^{\mathrm{ref}}$ for seed nodes
$k \in \known$ when reference positions exist. The gaussian kernel gives
least squares, and a robust kernel grows more slowly than the square for
large residuals and so limits the pull of an outlier, a range whose error far
exceeds the noise. This section generalizes~\eqref{eq:robust_argmin} into the
factor-graph form used throughout the paper.

\subsection{The Factor Graph}
\label{sec:factor_graph}

\begin{figure*}[!t]
  \centering
  \subfloat[fully meshed]{\includegraphics{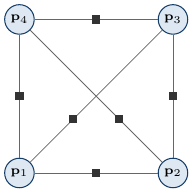}}\hfil
  \subfloat[multi-hop]{\includegraphics{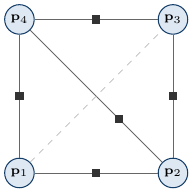}}\hfil
  \subfloat[mobile-assisted]{\includegraphics{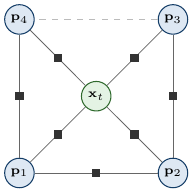}}\hfil
  \subfloat[mobile-only]{\includegraphics{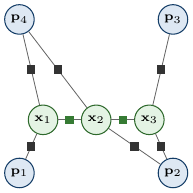}}\\[2pt]
  \includegraphics{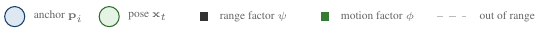}
  \caption{The four measurement topologies as factor graphs.}
  \label{fig:topologies_fg}
\end{figure*}

Split the nodes into anchors $\nodesA$ and mobile agents $\nodesM$, and the
unknowns into anchor positions and a discrete-time trajectory of $T$ poses,
$\allvars = \{\pos{i}\}_{i \in \nodesA} \cup \{\pose{t}\}_{t=1}^{T}$.
Edges split accordingly into anchor--anchor pairs $\edgesAA$ and
anchor--mobile pairs $\edgesAM$. Given all measurements $\mathcal{Z}$, with
errors assumed independent across pairs conditioned on $\allvars$, the
posterior factorizes over three factor types~\cite{dellaert_factor_2017},
drawn in the center panel of Fig.~\ref{fig:graphical_abstract} and,
without priors, per topology in Fig.~\ref{fig:topologies_fg}:
\begin{align}
\label{eq:factorization}
  &p(\allvars \mid \mathcal{Z}) \;\propto\;
  \prod_{(i,j) \in \edgesAA} \facRange{ij}(\pos{i}, \pos{j})
   \prod_{(i,t) \in \edgesAM} \facRange{it}(\pos{i}, \pose{t}) \nonumber\\
  &\quad\times
   \prod_{t=1}^{T-1} \facMotion{t}(\pose{t}, \pose{t+1})
   \prod_{k \in \known} \facPrior{k}(\pos{k})
   \prod_{t \in \knownM} \facPrior{t}(\pose{t}) .
\end{align}
A \emph{range factor} scores a residual through the kernel $\rhofun$, scaled
by a noise scale $\sigma$ (Section~\ref{sec:robust_loss}),
\begin{equation}
\label{eq:range_factor}
  \facRange{ij}(\pos{i}, \pos{j}) \;\propto\;
  \exp\!\left(-\rhofun\!\left(
    \frac{\meas{ij} - \|\pos{i} - \pos{j}\|}{\sigma}\right)\right),
\end{equation}
a \emph{motion factor} $\facMotion{t}$ encodes the motion model
between consecutive poses, which can be as weak as a Gaussian random walk
scaled by the maximum speed of the mobile when no odometry is
available~\cite{weber_automatisierte_2021}, and a \emph{unary prior}
$\facPrior{k}(\pos{k}) \propto
\exp(-\tfrac{1}{2}\|\pos{k}-\pos{k}^{\mathrm{ref}}\|^2_{\boldsymbol{\Sigma}_k})$
with covariance $\boldsymbol{\Sigma}_k$
carries the reference information~\cite{buchmayer_factor_2025}. A prior of
the same form, $\facPrior{t}(\pose{t})$ for $t \in \knownM$, applies to a
pose whose position an external reference supplies. The seed constraint of
\eqref{eq:robust_argmin} is the prior $\facPrior{k}$ in the limit
$\boldsymbol{\Sigma}_k \to 0$. A finite $\boldsymbol{\Sigma}_k$ encodes the
accuracy of a surveyed coordinate, so the soft prior covers exact and
approximate seeds. Maximum-a-posteriori (MAP) estimation over
\eqref{eq:factorization} recovers
\eqref{eq:robust_argmin}~\cite{grisetti_tutorial_2010}: taking the
negative logarithm gives
\begin{align}
\label{eq:map}
  \hat{\allvars}
  = {}&\arg\min_{\allvars}
  \sum_{(i,j) \in \edgesAA}
    \rhofun\!\left(\frac{r_{ij}(\allvars)}{\sigma}\right)
   + \sum_{(i,t) \in \edgesAM}
    \rhofun\!\left(\frac{r_{it}(\allvars)}{\sigma}\right) \nonumber\\
  &+ \sum_{t=1}^{T-1} c_{\phi}(\pose{t}, \pose{t+1})
   + \sum_{k \in \known} c_{\pi}(\pos{k})
   + \sum_{t \in \knownM} c_{\pi}(\pose{t}),
\end{align}
with range residuals $r_{ij}, r_{it}$ and the motion and prior cost terms
$c_{\phi}, c_{\pi}$. Equation~\eqref{eq:robust_argmin} is the special case
$\edgesAM = \emptyset$ with no motion factors, hard priors, and $\sigma$
absorbed into $\rhofun$.

Each topology of Section~\ref{sec:taxonomy} keeps a subset of these factors
(Fig.~\ref{fig:topologies_fg}). Every factor drawn there joins two variables,
so each panel shows the measurement graph of its topology with a factor node
on every measured pair. Panel~(d) adds the chain of motion factors between
consecutive poses, as in SLAM. The anchor-only topologies keep only the
anchor--anchor range factors and differ in the completeness of the mesh. The
mobile-assisted topology adds anchor--mobile range factors, the mobile-only
topology keeps only those, and motion factors are optional in both.

A static pair ranges in many epochs of a session, and
\eqref{eq:factorization} leaves open how they enter. The static batch engines
of Section~\ref{sec:algorithms} reduce them to one session median per pair.
The filters and the SLAM estimators take one factor per epoch, which treats
the epochs of a pair as independent given the positions. A bias that persists
per pair within a session, as in
UWB-ASC~\cite{schwarzbach_uwbasc_descriptor_2026}, breaks this
independence, yet a pair still weighs in proportion to its number of epochs,
and the repeats make a biased range look more certain without correcting it.
The engines therefore share the variables and the edge set of the graph while
their likelihoods differ, and each is compared here with its own data
interface. Range offsets and other primitives add nuisance states, which
Section~\ref{sec:nuisance} introduces once the identifiability of the
range-only graph is clear.

\subsection{Robust Losses and the Noise Scale}
\label{sec:robust_loss}

Each kernel weights a residual by $w(u) = \rhofun'(u)/u$, with
$u = r_{ij}/\sigma$ the scaled residual. The gaussian
kernel gives every residual unit weight, whereas a robust kernel lowers the
weight of large residuals~\cite{huber_robust_1964,zhang_parameter_1997}. Table~\ref{tab:kernels}
lists the three kernels of the reference implementations with their
weights. Their tuning constants~$c$ stay at the 95\,\%
Gaussian-efficiency values tabulated by Zhang~\cite{zhang_parameter_1997} for
the Huber kernel~\cite{huber_robust_1964} and the Cauchy kernel, not tuned
against an evaluation metric.

\begin{table}[!htbp]
  \caption{Kernels of the reference implementations and the linearized range
    factor.}
  \label{tab:kernels}
  \centering
  \footnotesize
  \setlength{\tabcolsep}{4pt}
  \begin{tabular}{@{}llll@{}}
    \toprule
    Kernel & $\rhofun(u)$ & $w(u) = \rhofun'(u)/u$ & $c$ \\
    \midrule
    gaussian & $u^{2}/2$ & $1$ & -- \\
    \addlinespace[2pt]
    huber & $\begin{cases} u^{2}/2, & |u| \le c\\
      c\,|u| - c^{2}/2, & |u| > c \end{cases}$ &
      $\min(1,\, c/|u|)$ & 1.345 \\
    \addlinespace[2pt]
    cauchy & $\tfrac{c^{2}}{2}\ln\bigl(1 + u^{2}/c^{2}\bigr)$ &
      $\bigl(1 + u^{2}/c^{2}\bigr)^{-1}$ & 2.385 \\
    \midrule
    Residual & \multicolumn{3}{l@{}}{$r_{ij} = \meas{ij} - \|\pos{i}-\pos{j}\|$,
      \ $u_{ij} = r_{ij}/\sigma$, with $\pose{t}$ for $\pos{j}$ on $\edgesAM$} \\
    Jacobian & \multicolumn{3}{l@{}}{$\partial\|\pos{i}-\pos{j}\|/\partial\pos{i}
      = -\partial\|\pos{i}-\pos{j}\|/\partial\pos{j} = \mathbf{e}_{ij}^{\top}$,} \\
    & \multicolumn{3}{l@{}}{$\mathbf{e}_{ij} = (\pos{i}-\pos{j})/\|\pos{i}-\pos{j}\|$,
      the blocks of a row of $\mathbf{J}$ in \eqref{eq:fim}} \\
    \bottomrule
  \end{tabular}
\end{table}

A kernel reads the residual in units of the noise scale $\sigma$, so $\sigma$
decides which ranges count as outliers. It should describe the spread of the
ranges that agree with the geometry, taken from a datasheet or estimated
from the ranges. A scale that outliers inflate, through a per-pair standard
deviation or a fit that NLOS ranges distort, weakens the kernel
(Section~\ref{sec:we_full_mesh}).

\begin{figure}[!htbp]
  \centering
  \subfloat[range residuals]{%
    \includegraphics{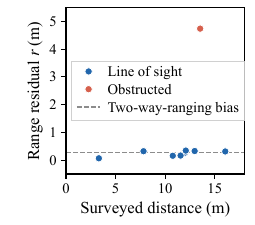}}\hfil
  \subfloat[cost per range]{%
    \includegraphics{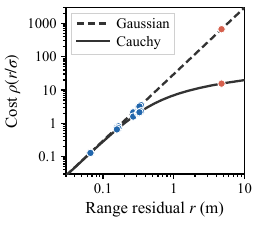}}\\[2pt]
  \subfloat[gaussian product]{%
    \includegraphics{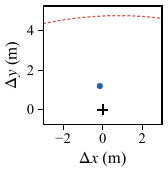}}\hfil
  \subfloat[cauchy product]{%
    \includegraphics{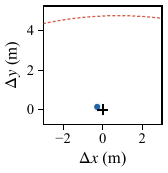}}\hfil
  \subfloat[gaussian mixture]{%
    \includegraphics{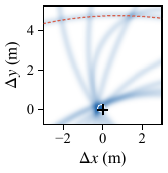}}\\[2pt]
  \includegraphics{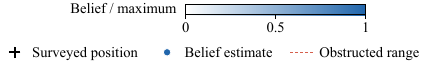}
  \caption{Ranges and position beliefs of one anchor at one epoch of the
    running example. The estimate dots cover the narrow product beliefs.}
  \label{fig:range_model}
\end{figure}

Fig.~\ref{fig:range_model} shows, on one anchor of the running example, how
the kernel decides the estimate when one range is an outlier. That range
belongs to the blocked pair, the only pair of the LOS deployment that an
obstruction biases by meters~\cite{schwarzbach_uwbasc_descriptor_2026}, and
its residual in panel~(a) lies far above those of the other eight ranges,
which sit near the two-way-ranging bias (Section~\ref{sec:we_bias}).
Panel~(b) gives the cost of each kernel. Panels~(c) and~(d) multiply the
likelihoods $e^{-\rhofun}$ of the nine ranges into the belief of the anchor,
the posterior of~\eqref{eq:factorization} for this anchor with the other
anchors held at their surveyed positions, and the dots mark the point
estimates that the grid filter of Section~\ref{sec:nonparametric} takes from a
belief. Under the gaussian kernel the blocked range pulls the belief toward
its arc, whereas the nearly flat cauchy cost of that range leaves the cauchy
belief where the other arcs meet. Section~\ref{sec:algorithms} shows the same
effect for the whole network.

Panel~(e) sums the nine gaussian likelihoods instead of multiplying them, as
the mixture variant of the grid filter does (Section~\ref{sec:nonparametric}),
and the sum also peaks where the other arcs meet, since the agreeing ranges
outvote the one displaced component of the blocked range.

\subsection{One Spectrum from Self-Calibration to SLAM}
\label{sec:spectrum}

The factorization also makes explicit how self-calibration, cooperative
localization, and radio SLAM relate. One graph already holds poses and
landmarks in factor-graph SLAM~\cite{dellaert_factor_2017}, and agents and
anchors in the message-passing treatments of cooperative
localization~\cite{wymeersch_cooperative_2009} and of multipath-based
SLAM~\cite{leitinger_scalable_2018}.
Joint formulations of calibration are older still. Moses et~al.\ located and
oriented sensor nodes from the arrival times and directions of calibration
signals emitted at unknown positions and derived the Cram\'er--Rao bound of
the estimates~\cite{moses_self-localization_2003}, and Taylor et~al.\
localized and calibrated a sensor network with a Bayesian filter while
tracking a moving target~\cite{taylor_simultaneous_2006}. Djugash et~al.\ ran
range-only SLAM with sensor-network nodes that also range to each
other~\cite{djugash_range-only_2006,djugash_motion-aided_2012}, and Meyer
et~al.\ combined the self-localization of a sensor network with the tracking
of targets by belief propagation, in
simulation~\cite{meyer_simultaneous_2012,meyer_distributed_2016}. Within the factor
graph the three problems differ in how the trajectory $\{\pose{t}\}$ is
treated~\cite{schwarzbach_terminology_2026,shi_anchor_2019,gentner_cooperative_2018}. Pure
self-calibration has no pose variables at all. Calibration with a helper
device introduces poses as nuisance variables, estimated jointly and then
discarded, because the mobile
exists to create informative range
factors~\cite{zhao_multiepoch_2025}. Radio SLAM estimates map and
trajectory jointly, with motion factors binding the poses into a
chain~\cite{gentner_simultaneous_2017,amjad_radio_2023}. This distinction
concerns the variables of interest and cuts across the two mobile topologies
of Section~\ref{sec:taxonomy}, which differ in the edge set alone.
Cooperative and network localization sit on the same spectrum and differ in
which nodes carry priors and which are mobile. Cooperative localization
spreads the measurements across peer
nodes~\cite{patwari_locating_2005,wymeersch_cooperative_2009},
and network localization estimates a static node set from a few priors or
none~\cite{priyantha_anchor-free_2003,shang_localization_2003,mao_wireless_2007}.
Fully cooperative settings with several mobiles add range factors on
mobile--mobile edges and one motion chain per agent, the same factor types
over a larger variable set. A multi-vehicle method has instead estimated an
anchor map per vehicle and merged the maps afterwards, trading the joint
posterior for tractability~\cite{weber_automatisierte_2021}. The presentation here stays
with the single-mobile case.

\textit{Pitfalls.}
\begin{itemize}
  \item A scale that the outliers inflate weakens a robust kernel, since
    $\sigma$ decides which ranges count as outliers.
  \item An engine that takes one factor per epoch weighs a biased pair by its
    epoch count, so the repeats make the bias look more certain without
    correcting it.
\end{itemize}

\section{What Ranges Can Identify}
\label{sec:identify}

Before any engine runs, the edge set and the priors decide whether the ranges
determine the geometry at all, and up to which transformation.

\subsection{Gauge and Alignment}
\label{sec:unary}

Range-only likelihoods depend on $\allvars$ only through pairwise distances,
so in the blind case the MAP objective \eqref{eq:map} inherits their
invariance under the Euclidean group, with three continuous degrees of freedom
plus a discrete reflection for $d=2$~\cite{burgess_toa_2015} and six plus a
reflection for $d=3$. A transformation of the unknowns that leaves every
predicted measurement unchanged is called a gauge freedom in bundle
adjustment~\cite{triggs_bundle_2000} and in visual-inertial
estimation~\cite{zhang_comparison_2018}, and in geodetic network adjustment it
appears as a datum defect~\cite{teunissen_zero_1985}. For ranges, the gauge is
this rigid transformation.

The reference-information regimes of Section~\ref{sec:taxonomy} reduce to the
number and placement of unary priors in \eqref{eq:factorization}. For $d = 2$
and non-collinear seeds, one prior fixes translation, a second fixes rotation
up to the reflection about the seed baseline, and a third resolves the
reflection. For $d=3$ a fourth seed off the plane of the others completes the
frame, so $d+1$ non-degenerate seeds fix it in $d$
dimensions~\cite{eren_rigidity_2004,aspnes_theory_2006}. In the
mobile-referenced regime the same count applies to the referenced poses in
$\knownM$. Partial priors, such as known mounting heights or anchors on a
known wall, fix only the coordinates they observe. In three dimensions,
anchors at nearly one height leave their vertical offsets weakly observable,
and nodes off the anchor plane restore them. Krape\v{z}
and Munih~\cite{krapez_anchor_2020} showed this with UWB modules on
the floor below the anchors.

Blind calibration ($\known = \knownM = \emptyset$) leaves the full Euclidean
invariance~\cite{kuang_microphone_2013}, so evaluation must align an estimate
to the reference frame before its errors are meaningful. The alignment
rotates and shifts the estimate, and mirrors it where needed, so that its
anchors lie as close as possible to the surveyed ones in the least-squares
sense. Umeyama's
construction~\cite{umeyama_least-squares_1991} computes it in closed form,
without the constraint against mirroring. The scale stays as estimated,
since the ranges fix it, so a scale error counts as an estimation error
rather than as gauge. The root-mean-square error (RMSE) of the aligned
anchor positions is the error measure of the examples.

An estimator still needs a frame to compute in, and three treatments of the
gauge give it one. A convention
fixes the frame through its defining nodes, a few nodes whose coordinates it
constrains. In the plane the first sits at the origin and the second on the
positive $x$-axis, which removes the continuous part of the gauge, and the
third at positive $y$ resolves the
reflection~\cite{hamer_self-calibrating_2018,loahavilai_rapid_2021,corbalan_self-localization_2023}.
In three dimensions the third lies in the $x$--$y$ plane and a fourth at
positive $z$. Every choice of defining nodes gives a frame of its own. Seeds
fix the reference frame instead, and a free gauge leaves the frame to the
solver until the alignment. A convention adds no information to the ranges,
so a batch estimator has the same minimizers in every such frame, up to the
rigid transformation and possible reflection between the
frames~\cite{triggs_bundle_2000,zhang_comparison_2018}.
Which of them an iterative solver reaches can still depend on the convention
through its steps, and on a flexible view (Section~\ref{sec:masking}) the end point does. The covariance
expressed in a convention frame vanishes at the first defining node and passes
the uncertainty of the defining nodes on to the other
nodes~\cite{triggs_bundle_2000}, whereas the aligned error of a given estimate
does not depend on the frame it is expressed in, which is why
Section~\ref{sec:crlb} states its bound after the alignment. An estimator that
places the other nodes from the defining ones in sequence, or a filter whose
prior is set in their frame, depends on the choice
(Section~\ref{sec:we_seeds}).

\subsection{Rigidity and Its Test}
\label{sec:rigidity}

\begin{figure}[!htbp]
  \centering
  \subfloat[flexible]{\includegraphics{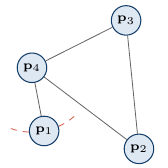}}\hfil
  \subfloat[rigid]{\includegraphics{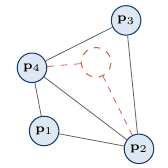}}\hfil
  \subfloat[globally rigid]{\includegraphics{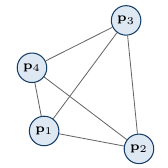}}\\
  \subfloat[static device]{\includegraphics{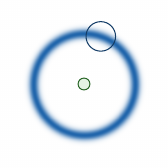}}\hfil
  \subfloat[straight walk]{\includegraphics{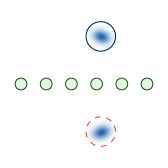}}\hfil
  \subfloat[walk with a turn]{\includegraphics{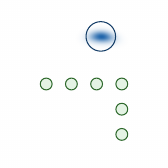}}
  \caption{Identifiability from exact ranges. (a)--(c) $\pos{1}$ with one to
    three ranges (lines). (d)--(f) Likelihood of one anchor (shading, true
    position circled) from six ranges at given poses (green). Dashed marks
    are other solutions.}
  \label{fig:identifiability}
\end{figure}

A graph of ranges at given node positions is rigid when no continuous motion
of the nodes other than a Euclidean one preserves all its ranges, and
globally rigid when its ranges determine the positions up to a Euclidean
transformation. It is generically rigid, or generically globally rigid, when
the property holds for almost every placement of its nodes, so that the edge
set alone decides it. Fig.~\ref{fig:identifiability}(a)--(c) draws the three
classes for one node. A complete anchor graph is globally rigid for any
configuration, and generically so is one that misses a single pair once it
holds at least $d+3$ anchors, so only the Euclidean invariance remains.

Thinning a mesh can add ambiguities at two thresholds of edge density, those
of rigidity and of global rigidity. Below the global-rigidity threshold, a
generically rigid graph admits finitely many incongruent embeddings whose
residuals are identical for exact ranges and nearly identical under
noise~\cite{hendrickson_conditions_1992,eren_rigidity_2004,aspnes_theory_2006}.
The multi-hop panel of Fig.~\ref{fig:topologies_fg} is such a graph, in which
$\pos{1}$ or $\pos{3}$ can flip across the $\pos{2}$--$\pos{4}$ diagonal.
Below the lower rigidity threshold, continuous flexes appear. Between the two
thresholds, an estimate can land on the wrong discrete branch even without
noise, a failure caused by topology rather than by the algorithm.

A rigidity test decides whether a graph clears the upper threshold. It needs
only the edges, so a deployment can run it on its planned or measured edge
set before any survey. In two dimensions a graph on at least four nodes is
generically globally rigid exactly when it is 3-connected and redundantly
rigid~\cite{jackson_connected_2005}, that is, when it stays connected after
the removal of any two nodes and stays rigid after the removal of any one
edge. Both conditions can be checked numerically. The rank of the rigidity
matrix, half the Jacobian of the squared ranges with respect to the node
coordinates~\cite{eren_rigidity_2004}, evaluated at random coordinates, equals
$2N-3$ for a generically rigid graph in the plane, and the same rank test
after the removal of each single edge checks redundant rigidity. A search
over the removals of node pairs for a disconnected remainder checks
3-connectivity. In three dimensions a randomized test decides generic global
rigidity through an equilibrium stress of maximal
rank~\cite{gortler_characterizing_2010}, a weighting of the edges whose forces
cancel at every node. The released toolkit implements the two-dimensional
test and evaluates the rank at small random perturbations of the surveyed
positions, which serve only as generic coordinates.

\subsection{Views of One Measurement Graph}
\label{sec:masking}

\begin{figure*}[!t]
  \centering
  \subfloat[fully meshed]{%
    \includegraphics{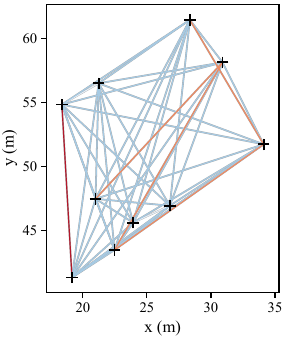}}\hfil
  \subfloat[multi-hop, $\kappa=3$]{%
    \includegraphics{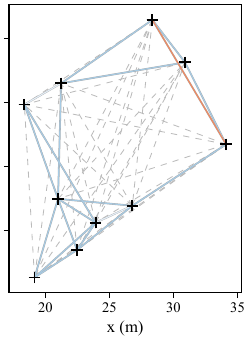}}\hfil
  \subfloat[mobile-assisted]{%
    \includegraphics{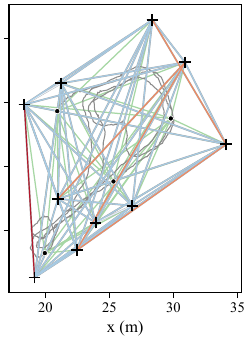}}\hfil
  \subfloat[mobile-only]{%
    \includegraphics{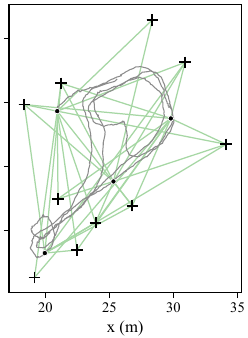}}\\[2pt]
  \includegraphics{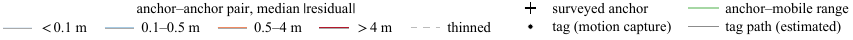}
  \caption{Views of the graph of session \texttt{s01} of the LOS deployment.
    Pair colors use the median absolute residual over all sessions of this
    deployment, and fans show the anchor--mobile ranges of four epochs.}
  \label{fig:views_graph}
\end{figure*}%
The examples vary the topology without recording a new campaign. The graph
of a session (Section~\ref{sec:records}) carries every edge of the session,
between anchors and, where a mobile node is present, between that node and the
anchors. An edge mask keeps a subset of these edges, optionally thins them,
and decides which anchors keep their surveyed prior. Its result is a
view~\cite{schwarzbach_uwbasc_descriptor_2026}. Within a session the views
differ in their edges and in the few priors they keep, which the blind
estimates of Section~\ref{sec:worked_examples} ignore, so the same ranges and
the same truth serve every topology.
Fig.~\ref{fig:views_graph} draws the four views on the walk \texttt{s01} of
the LOS deployment, and its tag path is the one that the cauchy factor-graph SLAM
estimator of Section~\ref{sec:algorithms} recovers on the mobile-only view.
The darkest pair color starts at \SI{4}{\meter}, the session-median threshold
of the outlier pairs of Section~\ref{sec:we_full_mesh}.

The multi-hop view thins the anchor mesh to each anchor's $\kappa$ nearest
neighbors by median range, emulating a network in which calibration
information must propagate over several
hops~\cite{corbalan_self-localization_2023}. The examples use $\kappa=3$, and
a sweep over $\kappa \in \{1,\dots,6,8\}$ moves the view across the thresholds
of Section~\ref{sec:rigidity}. On the running example the rigidity test puts the
threshold of generic global rigidity at $\kappa=\WEmhRigidKSetB$, in every
session of the LOS deployment, so the view at $\kappa=3$ is rigid but not
globally rigid. On the NLOS deployment the threshold lies at $\kappa=\WEmhRigidKSetA$, and the view at
$\kappa=3$ is flexible.

\begin{figure}[!t]
  \centering
  \includegraphics{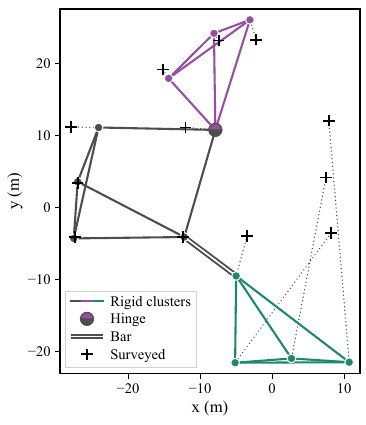}
  \caption{Cauchy factor-graph solution from the classical start on the
    flexible multi-hop view ($\kappa=3$) of session \texttt{s21} of the NLOS
    deployment, aligned on the lower gray triangle, with dotted lines to the
    surveyed positions.}
  \label{fig:anchor_scatter}
\end{figure}

Fig.~\ref{fig:anchor_scatter} shows how this view of the static session
\texttt{s21} of the NLOS deployment flexes. Its kept pairs form three rigid
clusters. Two share a single anchor, a hinge about which they can rotate
against each other, and the third hangs on a single pair, a bar, which fixes
only the distance between its two ends. The largest cluster is rigid but not
globally rigid, so one of its anchors can flip and carry the hinge along, and
the two smaller clusters can turn about the hinge and the bar, all without
changing a kept range. Started from the classical embedding of
Section~\ref{sec:algorithms}, the cauchy factor graph ends in such a
geometry, at an anchor RMSE of \SI{\WEmhFgCauchyKThreeSetA}{\meter}. Started
at the surveyed positions, it ends at \SI{\WEctlOracleKThreeSetA}{\meter} with
the same cost, so the kept ranges cannot tell the two geometries apart, and
the start decides.

\subsection{Motion in the Mobile Topologies}
\label{sec:motion}

In the mobile-only topology, only the shared poses couple the anchors. One
epoch leaves each anchor on a circle (a sphere for $d=3$) about the pose, and
enough epochs in general position, not all collinear and, for $d=3$, not all
coplanar, make the joint problem with at least $d+1$ anchors well posed up to
gauge~\cite{kuang_microphone_2013,zhao_multiepoch_2025}, as
Fig.~\ref{fig:identifiability}(d)--(f) shows for one anchor at given poses. A
static device does not provide this. Rotating each anchor independently about
a static tag changes the geometry by meters and leaves every range unchanged,
so the anchor geometry is not identifiable, and the running example uses the
mobile-only view on its walks only. Motion factors reduce the epochs
required. Ranges cannot distinguish chirality, so the reflection survives
unless the motion model or odometry carries signed turning information or
auxiliary state constraints, such as a floor bound on the anchor
heights~\cite{buchmayer_factor_2025}, suppress the mirrored solution. With
gyroscope heading changes in the motion model, mirrored anchor hypotheses
persisted along straight paths and were resolved at
turns~\cite{gentner_cooperative_2018}. In the
mobile-assisted topology the device adds range factors to an anchor mesh, and
it helps where the combined graph of anchors and device poses is rigid.

\subsection{The Cram\'er--Rao Bound as a Diagnostic}
\label{sec:crlb}

The examples use the Cram\'er--Rao bound to compare the conditioning of
views. For an anchor-only view with one range per observed pair and
zero-mean Gaussian errors of scale $\sigma$ in~\eqref{eq:measurement_model},
the Fisher information of the stacked anchor coordinates
$\mathbf{p} \in \mathbb{R}^{dN}$ of the $N$ anchors is
\begin{equation}
\label{eq:fim}
  \mathbf{F} \;=\; \frac{1}{\sigma^{2}}\,\mathbf{J}^{\top}\mathbf{J}
  \;+\; \sum_{k \in \known} \mathbf{E}_k \boldsymbol{\Sigma}_k^{-1} \mathbf{E}_k^{\top},
\end{equation}
where each row of the Jacobian $\mathbf{J}$, taken at the true geometry,
states how strongly the range of one pair reacts when an anchor coordinate
moves, and $\mathbf{E}_k$ selects the coordinates of seed~$k$. Up to a
scaling of its rows, $\mathbf{J}$ is the rigidity matrix of
Section~\ref{sec:rigidity} at the true geometry, so without priors the rank
of $\mathbf{F}$ is that of this matrix, and the rigid motions lie in its
nullspace. Where $\mathbf{F}$ is invertible, its inverse bounds the
covariance of any unbiased estimator from below, and for a blind
estimator, unbiased up to the gauge, its pseudo-inverse $\mathbf{F}^{+}$
bounds that of the aligned error to first
order~\cite{stoica_cramer-rao_1998,ash_relative_2008}. The square root of
$\operatorname{tr}(\mathbf{F}^{+})/N$ is the Gaussian bound. It limits
neither the error of a single estimate nor that of a biased estimator, and
on biased ranges it is no floor, since the per-pair bias breaks its model
(Section~\ref{sec:we_bias}). Without seeds the trace sums the reciprocal
nonzero eigenvalues of $\mathbf{F}$, so a near-flex mode with a small
eigenvalue can carry the bound. On the view of the running example thinned
to $\kappa=3$ one such mode contributes \SI{\WEcrlbMhKthreeWeakBthree}{\meter}
in quadrature to a bound of \SI{\WEcrlbMhKthreeBthree}{\meter}, and the
robust batch and Kalman estimators of Section~\ref{sec:algorithms} end below
it (Section~\ref{sec:we_thinned}).

\subsection{Other Measurement Primitives}
\label{sec:nuisance}

The range factor \eqref{eq:range_factor} assumes two-way ranges that measure
distance directly. Other primitives enter the same graph through their own
measurement models, which add nuisance states to $\allvars$
(Table~\ref{tab:nuisance_states}). Two-way ranging adds range offsets per
device~\cite{piavanini_calibration_2022,blueml_bias_2021}, one-way ToA adds
clock offsets and emission
times~\cite{mekonnen_self-calibration_2015,zhang_asynchronous_2024}, angles
add anchor orientations~\cite{lu_bayesian_2023}, as do phase differences of
arrival (PDoA)~\cite{ye_uwb_2025}, an array calibrated during operation adds
its gain and phase patterns~\cite{pohlmann_localization_2022}, RSS
adds the parameters of its path-loss model~\cite{patwari_locating_2005}, and
carrier phase adds ambiguities~\cite{hameed_lte_2023}. Each state changes what
a topology can identify. An unknown reference power $P_0$, the received power
at unit distance in the path-loss model, adds the scale to the gauge, because
a change of $P_0$ absorbs a common scaling of all distances. Bearings fix no
scale either, so their evaluation also aligns the scale.

\begin{table}[!htbp]
  \caption{Measurement models of the radio primitives and the nuisance
    states they add.}
  \label{tab:nuisance_states}
  \centering
  \footnotesize
  \setlength{\tabcolsep}{3pt}
  \begin{tabular}{@{}>{\raggedright}p{1.15cm}>{\raggedright}p{3.25cm}%
    >{\raggedright}p{1.45cm}>{\raggedright\arraybackslash}p{2.35cm}@{}}
    \toprule
    Primitive & Measurement model & Added states & Identifiability \\
    \midrule
    Two-way ranging &
      $\meas{ij} = \|\pos{i}-\pos{j}\|$\newline
      \hspace*{1.2em}$+\,b + \delta_i + \delta_j$ &
      offsets $b$, $\delta_i$ &
      $b$ correlates with scale; $b$ and the $\delta_i$ trade exactly,
      so the $\delta_i$ need a prior \\
    \addlinespace[2pt]
    One-way ToA, TDoA &
      $c_0\,\tilde{t}_{it} = \|\pos{i}-\pose{t}\|$\newline
      \hspace*{1.2em}$+\,c_0\,(\tau_t + \beta_i)$ &
      emission times $\tau_t$, clocks $\beta_i$ &
      a common clock offset is a gauge; each $\tau_t$ absorbs one arrival \\
    \addlinespace[2pt]
    AoA, PDoA &
      $\tilde{\theta}_{it} = \angle(\pose{t}-\pos{i}) - \alpha_i$ &
      orientations $\alpha_i$, array phases &
      bearings fix no scale; unknown $\alpha_i$ add rotation to the gauge,
      and each absorbs one bearing \\
    \addlinespace[2pt]
    RSS &
      $\tilde{P}_{it} = P_0$\newline
      \hspace*{1.2em}$-\,10\,\eta \log_{10}\|\pos{i}-\pose{t}\|$ &
      reference power $P_0$, path-loss exponent $\eta$ &
      unknown $P_0$ makes the scale a gauge for any $\eta$ \\
    \addlinespace[2pt]
    Carrier phase &
      $\lambda\tilde{\Phi}_{it} = \|\pos{i}-\pose{t}\|$\newline
      \hspace*{1.2em}$+\,\lambda\,(N_{i} + \gamma_i - \gamma_t)$ &
      integers $N_i$ per lock arc, phases $\gamma$ &
      $N_i$ and $\gamma_i$ trade exactly, so integer fixing needs calibrated
      phases \\
    \bottomrule
  \end{tabular}
  \par\smallskip
  \parbox{\linewidth}{\raggedright\footnotesize Noise is omitted. Tildes mark
    measurements, $c_0$ is the propagation speed, $\lambda$ the wavelength, and
    $\angle$ a bearing.}
\end{table}

The states also change how many anchors a topology needs. Per-device range
offsets need at least six anchors on a planar full mesh and eight in three
dimensions. One-way ToA or TDoA needs at least $d+2$
anchors on the mobile-only topology, since each emission time absorbs one
arrival, and unknown anchor clocks raise the number of epochs needed. The
examples use two-way ranges and add the offset states of the first
row in Section~\ref{sec:we_bias}. Standards supply parts of the graph. In
5G NR the network reports the coordinates and beam directions of its TRPs to
the location server~\cite{3gpp_ts_38305_2026}, which enter as priors on the
anchor poses, and the transmissions and measurements of one timing error
group, whose timing errors agree within a
margin~\cite{3gpp_ts_38305_2026,italiano_tutorial_2025}, share one offset state.

\textit{Pitfalls.}
\begin{itemize}
  \item A blind estimate carries an arbitrary rigid motion and reflection, so
    it is aligned before it is scored, and $d+1$ non-degenerate seeds,
    referenced poses, or an alignment to a map fix its frame
    (Section~\ref{sec:we_seeds}).
  \item Coordinates and covariances in the frame of a convention change with
    its defining nodes, so they are compared after alignment. An estimator
    that places nodes from the defining ones in sequence inherits the side of
    their line on which it puts each node and the errors of the ranges to them
    (Section~\ref{sec:we_seeds}).
  \item A connected view can still flip or flex, and with a static device its
    anchors can rotate about it. The rigidity test and the motion
    conditions reveal this before any survey.
  \item Read as a floor, the Gaussian bound misleads on biased ranges, whose
    bias breaks its model.
\end{itemize}

\section{Inference Engines}
\label{sec:algorithms}

\begin{table*}[!t]
  \caption{Reference implementations of the release by family and topology.}
  \label{tab:engines}
  \centering
  \footnotesize
  \setlength{\tabcolsep}{4pt}
  \begin{tabular}{@{}>{\raggedright}p{2.75cm}>{\raggedright}p{4.0cm}%
    >{\raggedright}p{3.2cm}>{\raggedright}p{4.3cm}%
    >{\raggedright\arraybackslash}p{2.0cm}@{}}
    \toprule
    Family & Fully meshed & Multi-hop & Mobile-assisted & Mobile-only \\
    \midrule
    MDS & \texttt{mds} & \texttt{mds} & \texttt{mds} & none \\
    \addlinespace[2pt]
    Batch nonlinear least squares &
      \texttt{factor\_graph}, \texttt{fg\_anchored}, \texttt{fg\_offset},
      \texttt{fg\_offset\_dev} &
      \texttt{factor\_graph}, \texttt{fg\_anchored} &
      \texttt{factor\_graph}, \texttt{fg\_offset}, \texttt{fg\_offset\_dev};
      moving device: \texttt{fg\_slam} &
      \texttt{fg\_slam} \\
    \addlinespace[2pt]
    Closed form (sequential trilateration) &
      \texttt{closed\_form} & \texttt{closed\_form} & \texttt{closed\_form} &
      none \\
    \addlinespace[2pt]
    Parametric recursive filtering & \texttt{ekf} & \texttt{ekf} &
      \texttt{ekf} & \texttt{ekf\_slam} \\
    \addlinespace[2pt]
    Nonparametric recursive filtering &
      \texttt{particle\_filter}, \texttt{grid\_filter} &
      \texttt{particle\_filter}, \texttt{grid\_filter} &
      \texttt{particle\_filter}, \texttt{grid\_filter} & none \\
    \bottomrule
  \end{tabular}
  \par\smallskip
  \parbox{\linewidth}{\raggedright\footnotesize Every method runs with the
    gaussian, huber, and cauchy kernels of Table~\ref{tab:kernels}, except
    \texttt{closed\_form}, which comes \texttt{faithful} or
    \texttt{disambiguated}. The variants of \texttt{mds} are
    \texttt{classical} and \texttt{smacof\_gaussian}, \texttt{smacof\_huber},
    and \texttt{smacof\_cauchy}. Among the filters, \texttt{ekf} adds
    \texttt{gated} and \texttt{grid\_filter} adds \texttt{mixture}.}
\end{table*}

Seen through the factor graph of Section~\ref{sec:formulation}, the
algorithm families are approximate ways to compute one posterior, either its
maximum or the belief itself, and each family is an inference engine, the
solver that computes this maximum or belief. Batch engines use all
measurements at once and reach a local optimum of~\eqref{eq:map}, whereas
recursive engines take them one at a time and approximate the belief. Each
engine reads the repeated ranges of a pair through its own data interface
(Section~\ref{sec:factor_graph}). Classical multidimensional scaling (MDS),
convex relaxations, and learned estimators solve surrogate problems instead.
This section is organized by engine, and Table~\ref{tab:engines} shows which
reference implementations of the release each family has on each topology.
The factor graph is the
representation that all engines share, and factor-graph optimization, the
factor graph of the examples, is the batch least-squares \mbox{engine applied to it.}
\onelinelonger

Learning enters the graph as front-ends that correct or reject ranges before a
classical engine~\cite{ridolfi_uwb_2021,sun_simultaneous_2025}, as regressors
that replace the estimator~\cite{yuan_large-scale_2025}, or as amortized
inference that maps measurements directly to device positions without anchor
variables~\cite{nguyen_uloc_2025}, as channel charting does up to a
deformation, with similarities of channel samples in place of range
factors~\cite{studer_channel_2018}. The release implements no learned engine,
and Section~\ref{sec:learning_eval} discusses how to evaluate one.

Each algorithm box states an engine as the release implements it, with the
method and variant names of the release in typewriter font, and its last line
names the module in \texttt{Code/methods} and its functions. The constants of
the boxes are untuned settings of the release. The stopping tolerances lie far
below the range noise, no measured session reaches the iteration caps, the
Levenberg--Marquardt iteration damps its steps on the schedule of Press
et~al.~\cite[Sec.~15.5]{press_numerical_2007}, and the extended Kalman filter
starts from a wide covariance.

\subsection{Multidimensional Scaling}

Classical MDS computes coordinates from a matrix of distances in one step.
Double centering turns the matrix of squared distances into a Gram matrix,
the inner products of the centered coordinates, whose leading $d$
eigenvectors, scaled by the square roots of their eigenvalues, give the
coordinates~\cite{torgerson_multidimensional_1952,gower_properties_1966}.
With noisy or NLOS distances the spectral step minimizes a misfit of the Gram
matrix, the strain, rather than solving~\eqref{eq:map}. It needs no
initialization, and its output is free in the gauge and aligned afterwards
(Section~\ref{sec:unary}). On the full mesh of the running example the
classical embedding of the session-median matrix lands at an aligned anchor
RMSE of \SI{\WEwtClassical}{\meter}. Following the steps of Shang
et~al.~\cite{shang_localization_2003}, which use classical MDS as the release
does, Krape\v{z} and Munih~\cite{krapez_anchor_2020} applied MDS to measured
UWB anchors. On their complete distance matrix it
outperformed semidefinite programming and iterative trilateration, and
semidefinite programming did better only in bias-free simulations with a
calibration plate of modules at known relative positions.

The algebra needs a complete matrix, so its native topology is the full mesh.
On an incomplete graph the MDS-MAP recipe of Shang
et~al.~\cite{shang_localization_2003} fills the missing entries with
shortest-path distances, which overestimate distances along sparse or curved
routes, the more so near the rigidity thresholds of
Section~\ref{sec:rigidity}. Scaling by majorizing a complicated
function (SMACOF) replaces the spectral step with iterative stress
majorization~\cite{deleeuw_applications_1977}. With unit weights, minimizing
the stress is the MAP problem~\eqref{eq:map} under the gaussian kernel on
$\edgesAA$ in the blind case, and robust kernels enter as reweighted stress
weights, so SMACOF links the spectral shortcut to the general optimizer of the
next subsection. Algorithm~\ref{alg:mds} gives both variants with the kernels
of Table~\ref{tab:kernels}.

\begin{algorithm}[!htbp]
  \caption{Classical MDS and robust SMACOF (\texttt{mds})}
  \label{alg:mds}
  \footnotesize
  \begin{algorithmic}[1]
    \Statex \textbf{Input:} session medians $\meas{ij}$ on the measured pairs
      $(i,j) \in \edges$ of a view with $N$ nodes, $\sigma$, $\rhofun$
    \State $D_{ij} \gets \meas{ij}$ on $\edges$, else the shortest-path
      length over $\edges$ (undefined if the view is disconnected)
    \State $\mathbf{G} \gets -\tfrac{1}{2}\,\mathbf{C}\,[D_{ij}^{2}]\,\mathbf{C}$
      with $\mathbf{C} = \mathbf{I} - \tfrac{1}{N}\mathbf{1}\mathbf{1}^{\top}$
    \State $\mathbf{G} = \mathbf{U}\boldsymbol{\Lambda}\mathbf{U}^{\top}$,
      with $\mathbf{U}_{d}$, $\boldsymbol{\Lambda}_{d}$ the $d$ leading
      eigenpairs
    \State $\mathbf{Y} \gets \mathbf{U}_{d}
      \max(\boldsymbol{\Lambda}_{d}, \mathbf{0})^{1/2}$, rows
      $\hat{\mathbf{p}}_{i}^{\top}$
      \label{alg:mds:classical}\Comment{classical ends here}
    \Repeat
      \State $\hat{d}_{ij} \gets \|\hat{\mathbf{p}}_{i} - \hat{\mathbf{p}}_{j}\|$,
        \ $u_{ij} \gets (\meas{ij} - \hat{d}_{ij})/\sigma$,
        \ $w_{ij} \gets w(u_{ij})$ on $\edges$, else $w_{ij} \gets 0$
      \State $V_{ij} \gets -w_{ij}$ and
        $B_{ij} \gets -w_{ij}\meas{ij}/\hat{d}_{ij}$ for $i \neq j$, with
        diagonals that make each row of $\mathbf{V}$ and $\mathbf{B}$ sum to zero
      \State $\mathbf{Y} \gets \mathbf{V}^{+}\mathbf{B}\,\mathbf{Y}$
        \Comment{Guttman transform}
    \Until{$\sum_{\edges}\rhofun(u_{ij})$, for the gaussian kernel the
      normalized stress, changes by less than $10^{-13}$ relative, or every
      coordinate moves by less than $10^{-12}$\,m, at most $10^{6}$ times}
    \Statex \textit{Release:} \texttt{mds.py} (\texttt{complete\_distances},
      \texttt{classical\_mds}, \texttt{smacof}, \texttt{smacof\_robust})
  \end{algorithmic}
\end{algorithm}

\subsection{Batch Nonlinear Least Squares}

Direct minimization of~\eqref{eq:map} by Gauss--Newton or
Levenberg--Marquardt (LM) iterations is the general engine of nonlinear least
squares. It assumes nothing about the factor structure, so the same solver
runs on all four topologies, and it stays tractable because its normal
equations inherit the sparsity of the factor
graph~\eqref{eq:factorization}~\cite{grisetti_tutorial_2010,dellaert_factor_2017}.
Robust kernels enter as iteratively reweighted least squares, each iteration
setting the weights of Section~\ref{sec:robust_loss} from the current
residuals. Range residuals make~\eqref{eq:map}
nonconvex, so the engine needs an initialization inside the basin of
attraction of the intended minimum, the set of starting points from which the
estimate converges to that minimum~\cite{yang_two-stage_2026,ortiz_uwb_2023}.
It also needs a gauge treatment (Section~\ref{sec:unary}), by a convention,
by unary priors, or by a free gauge whose nullspace the LM damping absorbs
before the estimate is aligned. Algorithm~\ref{alg:lm} gives the damped iteration of the release with
all three treatments, and convex relaxations remove the need for an
\mbox{initialization (Section~\ref{sec:convex}).}
\onelinelonger

\begin{algorithm}[!htbp]
  \caption{Robust LM of the factor-graph methods}
  \label{alg:lm}
  \footnotesize
  \begin{algorithmic}[1]
    \Statex \textbf{Input:} start $\boldsymbol{\theta}$ from
      line~\ref{alg:mds:classical} of Algorithm~\ref{alg:mds}
      (Section~\ref{sec:worked_examples}), the factors of the view, $\sigma$,
      $\rhofun$, and $f(\boldsymbol{\theta})$, the objective \eqref{eq:map}
      over these factors
    \State $\mu \gets 10^{-3}$
    \Repeat
      \State stack the residuals $r_{ij}$ of the range factors in $\mathbf{r}$
        and their Jacobian rows (Table~\ref{tab:kernels}) in $\mathbf{H}$,
        $\mathbf{W} \gets \operatorname{diag}\bigl(w(r_{ij}/\sigma)\bigr)$,
        all at $\boldsymbol{\theta}$
      \State $\mathbf{A} \gets \sigma^{-2}\mathbf{H}^{\top}\mathbf{W}\mathbf{H}
        + \sum_{k \in \known}\mathbf{E}_{k}\boldsymbol{\Sigma}_{k}^{-1}
        \mathbf{E}_{k}^{\top}$
      \State $\mathbf{b} \gets \sigma^{-2}\mathbf{H}^{\top}\mathbf{W}\mathbf{r}
        + \sum_{k \in \known}\mathbf{E}_{k}\boldsymbol{\Sigma}_{k}^{-1}
        (\pos{k}^{\mathrm{ref}} - \pos{k})$
      \Loop
        \State solve $(\mathbf{A} + \mu\operatorname{diag}\mathbf{A})\,
          \Delta\boldsymbol{\theta} = \mathbf{b}$
        \If{$f(\boldsymbol{\theta} + \Delta\boldsymbol{\theta})
          < f(\boldsymbol{\theta})$}
          \State $\boldsymbol{\theta} \gets \boldsymbol{\theta}
            + \Delta\boldsymbol{\theta}$, $\mu \gets \max(\mu/10, 10^{-12})$,
            and leave the loop
        \EndIf
        \State $\mu \gets 10\mu$, and return $\boldsymbol{\theta}$ if
          $\mu > 10^{12}$ \Comment{no downhill step}
      \EndLoop
    \Until{the accepted step lowers $f$ by at most $10^{-12}$ relative or
      moves every entry of $\boldsymbol{\theta}$ by less than $10^{-10}$\,m,
      at most $10^{4}$ times}
    \Statex \textit{Gauge:} \texttt{factor\_graph} uses the convention of
      Section~\ref{sec:unary} without its third node, $\pos{1} = \mathbf{0}$
      and the second coordinate of $\pos{2}$ at zero, so
      $\boldsymbol{\theta}$ holds $2N-3$ coordinates and $\known = \emptyset$.
      \texttt{fg\_anchored} keeps all $2N$ coordinates, first moves the start
      onto the seeds by the alignment of Section~\ref{sec:unary}, a
      shift for one seed, and leaves the directions that its priors do not
      fix to the damping. \texttt{fg\_slam} adds the poses to
      $\boldsymbol{\theta}$ and the random-walk factors of
      Section~\ref{sec:worked_examples}, with per-epoch step deviation~$q$, to $f$,
      $\mathbf{A}$, and $\mathbf{b}$, leaves the gauge to the damping, and
      stops within 100 iterations.
    \Statex \textit{Release:} \texttt{factor\_graph.py} (\texttt{solve\_nls},
      \texttt{solve\_nls\_anchored}), \texttt{fg\_slam.py}
      (\texttt{solve\_slam})
  \end{algorithmic}
\end{algorithm}

Because every iteration re-linearizes every factor, a later iteration can
repair a poor initialization, whereas the filter of the next subsection
commits to it at its first updates. On the running example, started from the
classical embedding, the blocked pair bends the geometry under the gaussian
kernel to an anchor RMSE of \SI{\WEfmBatchGaussBthree}{\meter}, with one
anchor \SI{\WEwtGaussMaxErr}{\meter} off. The cauchy kernel gives that pair a
weight of \WEwtCauchyObsWeight{} at the solution, so the other pairs set the
geometry, at \SI{\WEfmBatchCauchyBthree}{\meter} with no anchor farther than
\SI{\WEwtCauchyMaxErr}{\meter} off. The released
\texttt{factor\_graph} follows Dellaert and Kaess~\cite{dellaert_factor_2017}
with robust least squares as in Batstone et~al.~\cite{batstone_robust_2016}.
Its pose-chain variant \texttt{fg\_slam} runs on the mobile-only views and,
like \texttt{ekf\_slam}, starts from a classical embedding of the anchors and
every 25th tag pose, whose missing pairs shortest paths complete. The released
\texttt{closed\_form} places the anchors one after another by the sequential
trilateration of the author's prior
work~\cite{jung_auto-positioning_2022,jung_uncertainty_propagation_2025}, and
closed forms of this kind mostly serve as initializers.

\subsection{Parametric Recursive Filtering}

The parametric recursive filter keeps a Gaussian approximation of the
posterior and absorbs measurements one at a time. Motion factors drive the
prediction, and each range factor is linearized once, at the current mean, in
the update~\cite{kalman_new_1960,brown2012IntroductionRandomSignals}. Static
anchors enter the state with zero process noise, so calibration becomes
ordinary state estimation, and tag pose and anchor positions stacked in one
extended Kalman filter (EKF) give
EKF-SLAM~\cite{smith_estimating_1990,bailey_simultaneous_2006}. Robust kernels
act through the noise covariance rather than the cost. An M-estimation filter
in the sense of Huber~\cite{huber_robust_1964} inflates the variance of a
suspect range in inverse proportion to its kernel weight, and innovation
gating, the hard limit of the same idea, discards
updates whose normalized innovation fails a chi-square test.
Algorithm~\ref{alg:ekf} gives the scalar update of the release with both
mechanisms. The gate normalizes the innovation by its predicted variance, as
the chi-square test requires, whereas the weight scores it against $\sigma$
alone, as the batch reweighting scores a residual, so that one kernel scale
serves every engine. While the state covariance is still large, early updates
therefore read part of the state error as range error.

\begin{algorithm}[!htbp]
  \caption{Robust scalar EKF update (\texttt{ekf}, \texttt{ekf\_slam})}
  \label{alg:ekf}
  \footnotesize
  \begin{algorithmic}[1]
    \Statex \textbf{Input:} epoch stream of ranges $\meas{ij}$, $\sigma$,
      $\rhofun$ (gaussian for \texttt{gated}), start $\hat{\mathbf{p}}$ from
      line~\ref{alg:mds:classical} of Algorithm~\ref{alg:mds} (first 20
      epochs) in the gauge of Algorithm~\ref{alg:lm}
    \State $\boldsymbol{\Sigma} \gets (2\,\mathrm{m})^{2}\,\mathbf{I}$
      \Comment{static anchors, no prediction}
    \For{each epoch}
      \For{each pair $(i,j)$ ranged in the epoch, in the order of node IDs}
        \State $\nu \gets \meas{ij} - \|\hat{\mathbf{p}}_{i} - \hat{\mathbf{p}}_{j}\|$,
          \ $\mathbf{h}^{\top} \gets$ its Jacobian row
          (Table~\ref{tab:kernels}) at $\hat{\mathbf{p}}$
        \State \textbf{if} \texttt{gated} and $\nu^{2} > 9\,(\mathbf{h}^{\top}
          \boldsymbol{\Sigma}\mathbf{h} + \sigma^{2})$ \textbf{then} skip the pair
          \Comment{chi-square test of $\nu$}
        \State $w_{ij} \gets w(\nu/\sigma)$, and skip the pair if
          $w_{ij} < 10^{-12}$
          \Comment{one kernel scale with the batch engines}
        \State $s \gets \mathbf{h}^{\top}\boldsymbol{\Sigma}\mathbf{h}
          + \sigma^{2}/w_{ij}$, \ $\mathbf{k} \gets \boldsymbol{\Sigma}\mathbf{h}/s$
        \State $\hat{\mathbf{p}} \gets \hat{\mathbf{p}} + \mathbf{k}\,\nu$,
          \ $\boldsymbol{\Sigma} \gets \boldsymbol{\Sigma}
          - \mathbf{k}\,\mathbf{h}^{\top}\boldsymbol{\Sigma}$
      \EndFor
      \State $\boldsymbol{\Sigma} \gets (\boldsymbol{\Sigma}
        + \boldsymbol{\Sigma}^{\top})/2$
    \EndFor
    \Statex \texttt{ekf\_slam} holds the current tag position and all anchors
      in the state without a gauge constraint, adds $q^{2}\mathbf{I}$ to the
      covariance of the tag position between epochs, and processes the tag
      ranges of an epoch in anchor order.
    \Statex \textit{Release:} \texttt{ekf.py} and \texttt{ekf\_slam.py}
      (\texttt{run\_filter} in each)
  \end{algorithmic}
\end{algorithm}

Because the filter is causal, it suits calibration that must run online and
the mobile-only topology, where information arrives along the trajectory. The
price is the single linearization. At its first updates the filter commits to
the basin of attraction that contains its start, and the robust weights can
suppress precisely the innovations that would have pulled a bad start out, a
failure mode that may explain the lock-in on one mobile-only walk
(Section~\ref{sec:we_mobile}). Range-only SLAM treats this commitment as its
initialization problem and holds an anchor out of the state until its
candidate positions collapse~\cite{olson_robust_2006} or admits it at once as
a mixture of ring hypotheses~\cite{caballero_general_2010}.

\subsection{Nonparametric Recursive Filtering}
\label{sec:nonparametric}

The nonparametric filters represent the belief itself instead of a Gaussian
summary. A point-mass filter evaluates the posterior on a
grid~\cite{Bucy_digital_synthesis_non_linear_filters_1971,thrun2005probabilistic},
and a particle filter samples it~\cite{gordon_novel_1993}, so both keep
the multimodality of the range likelihoods that defeats a single
linearization. The cost grows exponentially with the state dimension, so one
belief per anchor replaces the joint posterior. Message passing, as in
nonparametric belief propagation~\cite{ihler_nonparametric_2005} and the
sum-product algorithm over a wireless network
(SPAWN)~\cite{wymeersch_cooperative_2009}, maps the inference onto the
network, so that each node combines messages from the nodes it ranges to,
which marginalize over their beliefs, and it needs no complete mesh. On the
mobile-only topology, filters that sample the trajectory and run one filter
per anchor~\cite{montemerlo_fastslam_2002,gentner_simultaneous_2017} and
grids corrected by incremental maximum-likelihood
steps~\cite{weber_automatisierte_2021} hold one belief per anchor given the
trajectory. Multipath-based radio SLAM, beyond the released engines, treats
each specular reflection as a range to a virtual anchor, the mirror image of
a physical anchor, and estimates the virtual anchors jointly with the
trajectory~\cite{gentner_multipath_2016,leitinger_scalable_2018}, which adds
detection and data association to the formulation of
Section~\ref{sec:formulation}.

The released implementations condition each anchor's belief on the current
estimates of the others and pass no messages, in one pass over the epoch
stream for the grid filter and two for the particle filter
(Algorithms~\ref{alg:pf} and~\ref{alg:grid}). Conditioning on point estimates
drops the uncertainty of the other anchors. On noise-free synthetic ranges
the release tests that an engine recovers the geometry where a view
permits it, SMACOF, the factor graph, and the EKF on every globally rigid
view, and the two filters only on the full anchor mesh. On views thinned to a
few neighbors the particle filter keeps, within its epoch budget, an error
that global rigidity does not cure, which the release attributes to this
conditioning without separating it from slow convergence. Thinning can also
drop the range between the first two defining nodes of the convention
the grid filter takes from the pipeline it ports, after which only the node
at the origin is placed. Both limits belong to the released
implementations.

\begin{algorithm}[!htbp]
  \caption{Per-anchor particle filter (\texttt{particle\_filter})}
  \label{alg:pf}
  \footnotesize
  \begin{algorithmic}[1]
    \Statex \textbf{Input:} epoch stream, $\sigma$, $\rhofun$, start
      $\hat{\mathbf{p}}_{i}$ from line~\ref{alg:mds:classical} of
      Algorithm~\ref{alg:mds} (first 20 epochs), node order and gauge of
      Algorithm~\ref{alg:lm}, $M = 2000$ particles per node, those of
      node~2 on the $x$-axis
    \For{sweep $1, 2$}
      \State draw $\mathbf{p}_{i}^{(m)} \sim \mathcal{N}\bigl(\hat{\mathbf{p}}_{i},
        (2\,\mathrm{m})^{2}\mathbf{I}\bigr)$, set
        $\omega_{i}^{(m)} \gets 1/M$, for all $i \ge 2$
      \For{each epoch}
        \For{$i = 2, \dots, N$}
          \State $\mathcal{R}_{i} \gets$ nodes $j$ that range to $i$ in the
            epoch, with $j < i$ in sweep~1 and $j \neq i$ in sweep~2
          \State $\omega_{i}^{(m)} \gets \omega_{i}^{(m)} \exp\bigl(
            -\sum_{j \in \mathcal{R}_{i}} \rhofun\bigl((\meas{ij}
            - \|\mathbf{p}_{i}^{(m)} - \hat{\mathbf{p}}_{j}\|)/\sigma\bigr)\bigr)$,
            normalized over $m$
          \If{$1/\sum_{m}(\omega_{i}^{(m)})^{2} < M/2$}
            \State resample systematically, add $\mathcal{N}\bigl(\mathbf{0},
              (0.02\,\mathrm{m})^{2}\mathbf{I}\bigr)$ jitter, and set
              $\omega_{i}^{(m)} \gets 1/M$
          \EndIf
          \State $\hat{\mathbf{p}}_{i} \gets \sum_{m}\omega_{i}^{(m)}
            \mathbf{p}_{i}^{(m)}$ \Comment{used at once for later $i$}
        \EndFor
      \EndFor
    \EndFor
    \Statex \textit{Release:} \texttt{particle\_filter.py}
      (\texttt{particle\_filter\_anchors})
  \end{algorithmic}
\end{algorithm}

\begin{algorithm}[!htbp]
  \caption{Per-anchor grid filter (\texttt{grid\_filter})}
  \label{alg:grid}
  \footnotesize
  \begin{algorithmic}[1]
    \Statex \textbf{Input:} every $\max(1, \lfloor T/75 \rfloor)$-th of the
      $T$ epochs, $\sigma$, $\rhofun$, node order and gauge of
      Algorithm~\ref{alg:lm}
    \Statex \textbf{Grids:} cells $\mathbf{g}$ of 0.25\,m on a square of
      half-width $\ell$, the largest median range from node~1 plus 5\,m, and
      cells $x \in [0, \ell]$ for node~2
    \Statex \textbf{Start:}
      ${x_{i} = (\hat{x}_{2}^{2} + \meas{i1}^{2} - \meas{i2}^{2})/(2\hat{x}_{2})}$,
      ${y_{i} = +(\meas{i1}^{2} - x_{i}^{2})^{1/2}}$, the trilateration of
      node $i \ge 3$ from nodes~1 and~2
    \State $\mathrm{bel}_{i} \gets$ uniform for $i \ge 2$, node~1 at the origin
    \For{each epoch}
      \State \textbf{if} $\meas{12}$ is ranged \textbf{then}
        $\mathrm{bel}_{2}(x) \gets \mathrm{bel}_{2}(x)
        \exp\bigl(-\rhofun((\meas{12} - x)/\sigma)\bigr)$ and
        $\hat{x}_{2} \gets \arg\max \mathrm{bel}_{2}$
        \label{alg:grid:hist}
      \For{$i = 3, \dots, N$}
        \State $\mathcal{R}_{i} \gets$ nodes $j \neq i$ with an estimate and a
          range $\meas{ij}$ in the epoch
        \If{$\mathrm{bel}_{i}$ has not started}
          \State multiply $\mathrm{bel}_{i}$ by $\mathcal{N}\bigl(\mathbf{g};
            (x_{i}, y_{i})^{\top}, (0.5\,\mathrm{m})^{2}\mathbf{I}\bigr)$ if
            $\hat{x}_{2} > 0$, $\meas{i1}$ and $\meas{i2}$ are ranged, and
            $\meas{i1} > |x_{i}|$
          \State otherwise keep it uniform if $|\mathcal{R}_{i}| \ge 3$, else
            skip node~$i$
        \EndIf
        \If{$\mathcal{R}_{i} \neq \emptyset$}
          \State $u_{ij}(\mathbf{g}) \gets (\meas{ij} - \|\mathbf{g}
            - \hat{\mathbf{p}}_{j}\|)/\sigma$ for $j \in \mathcal{R}_{i}$ and
            every cell $\mathbf{g}$
          \State $\mathrm{bel}_{i}(\mathbf{g}) \gets \mathrm{bel}_{i}(\mathbf{g})
            \prod_{j \in \mathcal{R}_{i}}
            \exp\bigl(-\rhofun(u_{ij}(\mathbf{g}))\bigr)$
            \label{alg:grid:product}
          \State normalize, raise every cell to at least $6 \cdot 10^{-3}$ of
            the uniform value, normalize \label{alg:grid:floor}
        \EndIf
        \State $\hat{\mathbf{p}}_{i} \gets$ mean of the cells within 4\,m of
          $\arg\max \mathrm{bel}_{i}$, weighted by $\mathrm{bel}_{i}$
      \EndFor
    \EndFor
    \Statex Line~\ref{alg:grid:hist} normalizes and raises like
      line~\ref{alg:grid:floor}. \texttt{mixture} replaces the product of
      line~\ref{alg:grid:product} by
      $\sum_{j \in \mathcal{R}_{i}} \exp\bigl(-u_{ij}(\mathbf{g})^{2}/2\bigr)$
      and takes the gaussian kernel on line~\ref{alg:grid:hist}.
    \Statex \textit{Release:} \texttt{grid\_filter.py}
      (\texttt{run\_grid\_filter})
  \end{algorithmic}
\end{algorithm}

That pipeline, the probability grid of the author's prior
work~\cite{jung_auto-positioning_2022,jung_uncertainty_propagation_2025}, is
one instance of this engine. Its original implementation multiplies the grid
at each epoch by the sum of the range likelihoods of the reference anchors
rather than by their product. Normalized, the sum is a linear opinion
pool~\cite{genest_combining_1986} and the exact posterior of a model in which
one range of the epoch, picked with equal probability, is informative and the
others are outliers uniform over the measurable ranges.
The later version~\cite{jung_uncertainty_propagation_2025} multiplies over the
references, each likelihood averaged over the most probable cells of the
reference grid. The released
\texttt{grid\_filter} multiplies the range factors~\eqref{eq:range_factor} at
the reference estimates and keeps the original sum as its \texttt{mixture}
variant, which Fig.~\ref{fig:range_model}(e) draws.

\subsection{Convex Relaxation and Certifiable Estimation}
\label{sec:convex}

Convex relaxations replace the nonconvex problem~\eqref{eq:robust_argmin} by
a semidefinite program over the Gram matrix of the positions or over the
matrix of squared distances, without its rank
constraint~\cite{biswas_semidefinite_2004,dokmanic_euclidean_2015}, and need
no initialization. With exact distances and anchors, the semidefinite
relaxation of sensor network localization is exact precisely when the
instance is uniquely localizable~\cite{so_theory_2007}, a condition stronger
than the global rigidity of Section~\ref{sec:rigidity}, since it also excludes
realizations in higher dimensions. Under noise the relaxed solution
is rounded to rank~$d$ and refined locally, and a dual certificate shows when
the refined estimate is the global optimum of the nonconvex problem, as in
certifiably correct range-aided SLAM~\cite{papalia_certifiably_2024}. The
semidefinite variables grow with the square of the number of nodes, and
robust losses enter through graduated non-convexity, which wraps such solvers
instead of relaxing the robust problem~\cite{yang_graduated_2020}. On biased
ranges a minimum of lower cost can be the worse geometry, as on randomly
thinned views of the NLOS deployment (Section~\ref{sec:we_thinned}), so a
certificate of global optimality does not certify accuracy. The release
implements none, and the decision tree offers the family at its multi-hop
leaf (Section~\ref{sec:decision_tree}).

\textit{Pitfalls.}
\begin{itemize}
  \item A recursive filter commits at its first updates to the basin of
    attraction that contains its start, so a poor start can persist, whereas a
    batch engine re-linearizes every factor in every iteration and can leave
    it.
  \item Classical MDS needs no start but solves a surrogate problem, and on a
    thinned mesh its shortest-path completion overestimates the missing
    distances.
\end{itemize}

\section{Calibration in Practice: Lessons Learned}
\label{sec:worked_examples}

\begin{figure*}[!t]
  \centering
  \subfloat[LOS deployment]{%
    \includegraphics{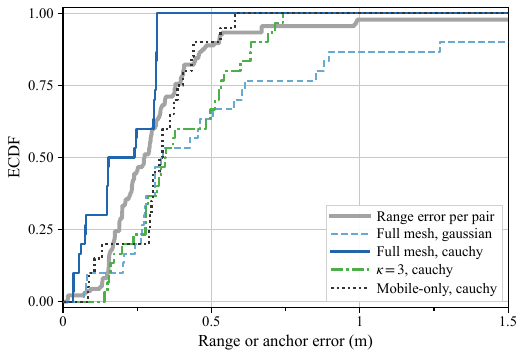}}\hfil
  \subfloat[NLOS deployment]{%
    \includegraphics{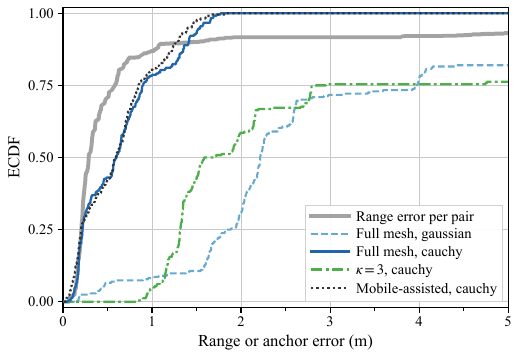}}
  \caption{Empirical cumulative distribution functions (ECDF) of the absolute
    session-median range residual per anchor pair and of the aligned error per
    anchor of the factor graph, pooled over the sessions. Each panel clips its
    own abscissa.}
  \label{fig:error_propagation}
\end{figure*}%

The examples run the reference implementations of
Section~\ref{sec:algorithms} on the views of Section~\ref{sec:masking} under
one protocol. Each result belongs to one cell, a combination of session, view,
method, and variant. Every estimate except the seeded ones of
Section~\ref{sec:we_seeds} is blind, and each is scored by the aligned anchor RMSE of
Section~\ref{sec:unary} against the surveyed coordinates. On the mobile-only
view the absolute trajectory error and the relative pose
error~\cite{sturm_benchmark_2012} also score the
recovered path, at the epochs with motion-capture truth. The iterative engines
start from a classical MDS embedding, so that the comparison concerns
estimation rather than initialization~\cite{yang_two-stage_2026}. The grid
filter keeps the closed-form start of the pipeline it ports, and both SLAM
estimators use a random-walk motion prior whose per-epoch step has a standard
deviation of \SI{1.5}{\meter\per\second} times the median epoch spacing. Every kernel reads
the residuals in units of one noise scale per session, the larger of the median
per-pair standard deviation of the ranges and 1.4826 times the median absolute
deviation of the residuals of a classical MDS embedding of the session medians.
It is set on the fully meshed view and kept for every view of the session, so
that the noise model cannot absorb the connectivity effect.

Aggregates over the sessions of a deployment are medians, with particle-filter
cells as means over \WEpfSeeds{} seeds. Three settings were fixed after results
had been inspected. The shared noise scale replaced per-method scales, the EKF
and the particle filter start from the MDS embedding because a closed-form
start trapped the robust kernels on the NLOS deployment, and the particle
filter keeps a second sweep, which lowered its robust errors there. Each
iterative family first passed the recovery test of
Section~\ref{sec:nonparametric}.

The examples cover two-dimensional UWB two-way ranging at the two deployments
of UWB-ASC and one public scan, and they handle NLOS links only through
kernels, gates, and mixtures, without channel-based
identification~\cite{ridolfi_uwb_2021}. The sessions of a deployment repeat one
anchor network, \WEnodeSets{} anchor configurations in all, and the neighbor,
random, and noise-scale sweeps run on one static session per deployment, so
the lessons are demonstrations within that scope.

\phantomsection\label{sec:we_synthetic}%
Synthetic ranges check the lessons where measured ranges cannot split their
error into noise, NLOS excess, and bias. The range simulator of
UWB-ASC~\cite{schwarzbach_uwbasc_descriptor_2026}, a seeded port of the
author's range-error model~\cite{schwarzbach_statistical_2022}, adds to
Gaussian hardware noise of \SI{\WEsynSigmaHw}{\meter} either per-range outliers
and an NLOS excess drawn once per pair, as on the measured pairs biased by
meters (Section~\ref{sec:we_full_mesh}), or a common bias equal to the
per-pair median of the LOS deployment (Section~\ref{sec:we_bias}), in \WEsynDraws{} draws
each on the anchors of the running example.

On the NLOS deployment the outlier pairs shift the whole anchor-error curve of
the gaussian factor graph by meters (Fig.~\ref{fig:error_propagation}), the
cauchy kernel removes most of that shift (Section~\ref{sec:we_full_mesh}), and
under the same kernel the flexible $\kappa=3$ view ends at larger errors, since
its start decides the minimum (Section~\ref{sec:we_thinned}).
On the LOS deployment the full-mesh cauchy anchors stay at or below the size of
the range bias, which no kernel removes (Section~\ref{sec:we_bias}), and the
other curves lie to their right. Table~\ref{tab:lessons_effect} gives the anchor
error without and with the action of each lesson and finding.

\begin{table}[!htbp]
  \caption{Anchor RMSE without and with the action of each finding.}
  \label{tab:lessons_effect}
  \centering
  \footnotesize
  \setlength{\tabcolsep}{3pt}
  \begin{tabular}{@{}>{\raggedright}p{3.9cm}>{\raggedright}p{2.6cm}rr@{}}
    \toprule
    & & \multicolumn{2}{c}{RMSE (m)} \\
    \cmidrule(l){3-4}
    Finding and action & Where & without & with \\
    \midrule
    Loss before engine (\ref{sec:we_full_mesh}): gaussian $\to$ cauchy & full meshes, NLOS deployment & 3.94 & 0.79 \\
    \addlinespace[2pt]
    Start decides (\ref{sec:we_thinned}): classical start $\to$ oracle start & globally rigid random views of \texttt{s03} & 4.03 & 0.24 \\
    \addlinespace[2pt]
    Range bias (\ref{sec:we_bias}): no offsets $\to$ per-device offsets & full meshes, LOS deployment & 0.23 & 0.13 \\
    \addlinespace[2pt]
    Mobile-only (\ref{sec:we_mobile}): \texttt{ekf\_slam} $\to$ \texttt{fg\_slam} & walk \texttt{s01}, LOS deployment & 1.88 & 0.36 \\
    \addlinespace[2pt]
    Seeds (\ref{sec:we_seeds}): blind $\to$ three seeds & $\kappa=3$ view of \texttt{s03} & 0.43 & 0.71 \\
    \bottomrule
  \end{tabular}
  \par\smallskip
  \parbox{\linewidth}{\raggedright\footnotesize Aligned anchor RMSE of the
    factor graph or of the engines a row names, under the cauchy kernel unless
    the action changes it, as medians where a row covers several sessions or
    views. The oracle start is the survey.}
\end{table}

\subsection{Full Meshes: The Loss Before the Engine}
\label{sec:we_full_mesh}

\begin{figure*}[!t]
  \centering
  \subfloat[LOS deployment, \texttt{s03}]{%
    \includegraphics{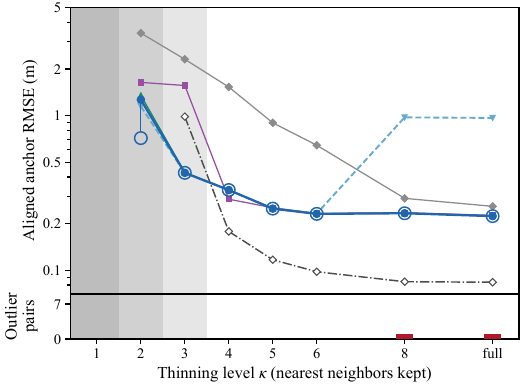}}\hfil
  \subfloat[NLOS deployment, \texttt{s21}]{%
    \includegraphics{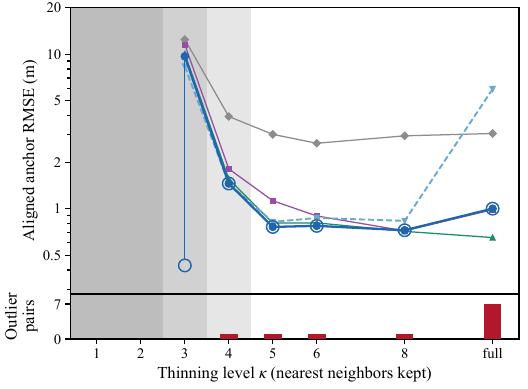}}\\[2pt]
  \includegraphics{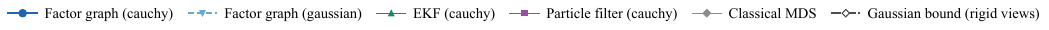}
  \caption{Anchor RMSE against the thinning level $\kappa$. Bands shade
    rigid but not globally rigid (light), flexible (medium), and
    disconnected (dark) views. On the shaded views, rings mark the cauchy
    factor graph started from the survey, an oracle start, with a stem to its
    classical-start dot where the two differ. Bars count the kept outlier
    pairs. The particle filter conditions each anchor on the point estimates
    of the others (Section~\ref{sec:nonparametric}).}
  \label{fig:multi_hop_ksweep}
\end{figure*}

On a full mesh, choose a robust loss and set its scale first, and calibrate
with the engine of the positioning system, since the engines then reach
similar errors. NLOS propagation biases several pairs of the NLOS deployment
by meters~\cite{schwarzbach_uwbasc_descriptor_2026}. Under a quadratic loss the
outlier pairs, those whose median residual in a session exceeds
\SI{4}{\meter}, bend the whole geometry, and the cauchy kernel sets them aside
(Table~\ref{tab:lessons_effect}). Under that kernel the medians of the engines
span \WEfmCauchySpanSetA\,m, against \WEfmGaussSpanSetA\,m under the gaussian
kernel, and the EKF and the grid filter even end below the batch solver in
almost every full-mesh and mobile-assisted session of that deployment,
possibly because they read the epoch stream rather than one median per pair
(Section~\ref{sec:factor_graph}).
On the meshes of the LOS deployment the
engines end within \SI{\WEfmEngineSpreadSetB}{\meter} of each other.

Setting the scale is the proviso, because the scale decides whether the kernel
can set the outliers apart. On the static session \texttt{s21}, NLOS
distortion inflates $\sigma$ to
\SI{\WEsigMdsAtwentyone}{\meter}, against \SI{\WEsigMdsBthree}{\meter} on the
running example, and a quarter of it brings every cauchy engine to
\WEctlSweepQuarterSpanSetA\,m. The per-range outliers of the synthetic NLOS
ranges inflate $\sigma$ through its per-pair standard deviation, and at that
scale the cauchy kernel gains little, whereas with its scale at the hardware
noise, a datasheet value in the sense of Section~\ref{sec:robust_loss}, it
removes most of the error.
The public scan shows no gain from the robust loss
(Section~\ref{sec:we_thinned}), so the lesson rests on the two deployments
and, for the loss, on the synthetic ranges.

\subsection{Thinned and Partial Meshes: The Start Decides}
\label{sec:we_thinned}

On thinned or partial meshes, test the result at surveyed check points,
because the start decides which minimum a solver reaches. On the
nearest-neighbor views of Section~\ref{sec:masking}, the robust batch and
Kalman estimators keep the correct branch while a view stays rigid, at
\WEmhRobustKThreeSpanSetB\,m on \texttt{s03} at $\kappa=3$, and fail once it
turns flexible (Fig.~\ref{fig:multi_hop_ksweep}), where the start decides, as
in Fig.~\ref{fig:anchor_scatter}. Classical MDS fails metrically on the
$\kappa=3$ views, since shortest-path completion overestimates every missing
distance. On \texttt{s21} the
inflated links look long and go first, so thinning keeps fewer outlier pairs
and the gaussian factor graph improves from the full mesh to $\kappa=8$, while
on \texttt{s03} thinning below $\kappa=8$ removes the blocked pair.

Global rigidity is necessary for a unique geometry but not sufficient for a
batch solver to find it. Random thinning, which draws each node's $\kappa$
neighbors at random, mostly gives globally rigid views. From the classical
start the factor graph ends far from the survey on \WEctlRandSplitSetB{} of the
\WEctlRandRigidSetB{} globally rigid views of \texttt{s03}, while a start from
the survey, an oracle, ends close to it on all of them
(Table~\ref{tab:lessons_effect}), at the lower robust cost wherever the two
starts disagree. On \texttt{s21} the start of
lower cost would have picked the worse geometry on \WEctlRandCostWrongSetA{} of
the \WEctlRandSplitSetA{} views where the starts disagree, so the cost cannot
tell which start to trust. Van~Herbruggen et~al.\ found the trilateration
start of a seeded multi-hop UWB network to vary strongly with the choice of the
seeds~\cite{herbruggen_multihop_2023}. Even on a full mesh,
the classical start led the cauchy factor graph at the hardware scale to a
minimum of higher cost than a start at the true geometry in
\WEsynNlosStartWorse{} of the \WEsynDraws{} synthetic NLOS draws.

The partial mesh of the public TWR-CLOUD-Bern scan
(Section~\ref{sec:running_example}) shows the same on a third layout, where
every engine ends \WEextSharedSpan\,m from the survey from the classical start
and the factor graph started from the survey \WEextSurveyedSpan\,m, with the
three kernels within \SI{\WEextKernelDiff}{\meter} of each other, so the scan
shows no gain from the robust loss.

\subsection{Range Bias and Offset States}
\label{sec:we_bias}

\begin{figure}[!htbp]
  \centering
  \includegraphics{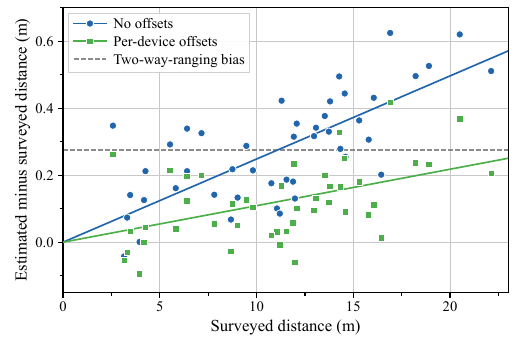}
  \caption{Estimated minus surveyed distance of every anchor pair on the full
    mesh of \texttt{s03} of the LOS deployment, cauchy factor graph without and
    with per-device offsets. Solid lines mark the expansion of each estimate.}
  \label{fig:pair_stretch}
\end{figure}

On a LOS deployment, estimate range offsets, which lower the error floor that
a systematic ranging bias sets and a robust loss cannot remove. On the LOS
deployment the anchor-to-anchor ranges exceed the surveyed distances by a
per-pair median of
\SI{+\WEresAaMedSetB}{\meter}~\cite{schwarzbach_uwbasc_descriptor_2026}, close
to the median offset of \SI{0.33}{\meter} that Piavanini
et~al.~\cite{piavanini_calibration_2022} measured between anchors before
correcting the antenna delays, which are specific to each
device~\cite{yudanto_device-level_2023}. On the meshes of this deployment the
blind cauchy estimates end at medians of
\WEfmCauchyFloorSetB\,m whatever the engine. The bias lengthens every pair, and
the blind estimate absorbs it by spreading the anchors apart
(Fig.~\ref{fig:pair_stretch}), an expansion of \WEscaleExpandSpanSetB\,\% in a
similarity fit to the survey that carries \WEscaleShareSpanSetB\,\% of the
mean~squared~error.

The offset variants of the factor graph fit the first row of
Table~\ref{tab:nuisance_states}, \texttt{fg\_offset} with the common offset $b$
and \texttt{fg\_offset\_dev} also with one offset $\delta_i$ per device under a
zero-mean Gaussian prior of width \SI{\WEoffWidth}{\meter}, fixed before the
run. The common offset lowers the median to \SI{\WEoffFmCommonSetB}{\meter},
per-device offsets lower it further (Table~\ref{tab:lessons_effect}), and an
expansion of \SI{\WEscaleExpandDeviceSetB}{\percent} remains. On the synthetic
ranges the common offset recovers a purely common bias in full, so the
expansion that remains points to the part of the measured bias that varies by
pair. One surveyed distance checks the
scale. On the NLOS deployment the offsets help only at a reduced $\sigma$.

\subsection{The Mobile Topologies}
\label{sec:we_mobile}

\begin{figure}[!htbp]
  \centering
  \subfloat[\texttt{s01}, anchor positions]{%
    \includegraphics{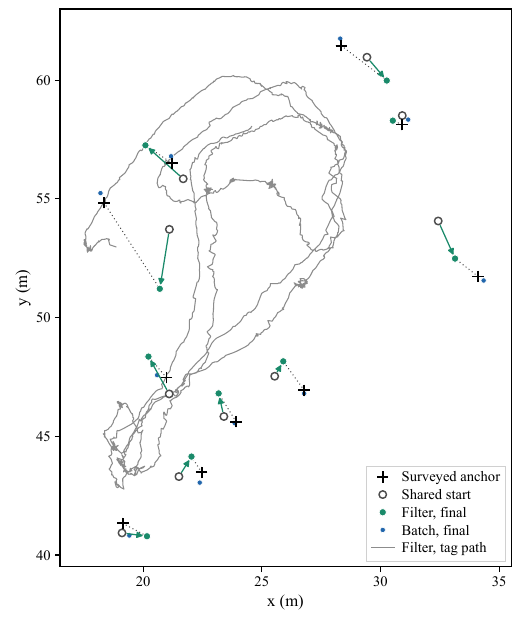}}\\
  \subfloat[\texttt{s01}, anchor error]{%
    \includegraphics{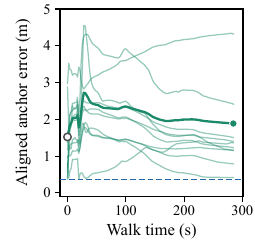}}\hfill
  \subfloat[\texttt{s02}, anchor error]{%
    \includegraphics{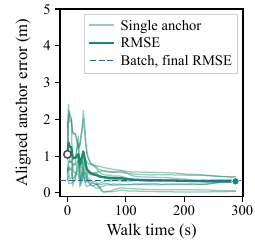}}
  \caption{Causal filter (\texttt{ekf\_slam}) and batch solver
    (\texttt{fg\_slam}), both cauchy, on the mobile-only view of the LOS
    deployment. Arrows
    lead from the shared start to the final filter anchors, dotted lines from
    these to the survey.}
  \label{fig:slam_trajectory}
\end{figure}

A static tag on its surveyed marker adds little to the mesh of the NLOS
deployment (Fig.~\ref{fig:error_propagation}(b)), and mostly to the gaussian
estimators. On the mobile-only view of the walks of the LOS deployment the
robust loss helps on the anchors and on the recovered path alike, while
observing the anchors only through the moving tag costs accuracy
(Fig.~\ref{fig:error_propagation}(a)), although each walk holds about
\WEslamEpochsK{} thousand tag-ranging epochs. Unlike on the meshes, the engine
matters here. From the shared bipartite classical-MDS start
(Fig.~\ref{fig:slam_trajectory}(a)) the batch solver re-linearizes its way out
on both walks, whereas the causal filter commits to the basin of attraction
that contains that start, as Section~\ref{sec:algorithms} anticipates. It locks in on
\texttt{s01} (Fig.~\ref{fig:slam_trajectory}(b),
Table~\ref{tab:lessons_effect}) and ends below the batch solver on \texttt{s02}
(Fig.~\ref{fig:slam_trajectory}(c)). On this topology batch re-linearization is
the safer choice, even for a deployment that tracks with a filter.

\subsection{Defining Nodes, Seeds, and the Absolute Frame}
\label{sec:we_seeds}

\begin{figure*}[t]
  \centering
  \includegraphics[width=\textwidth]{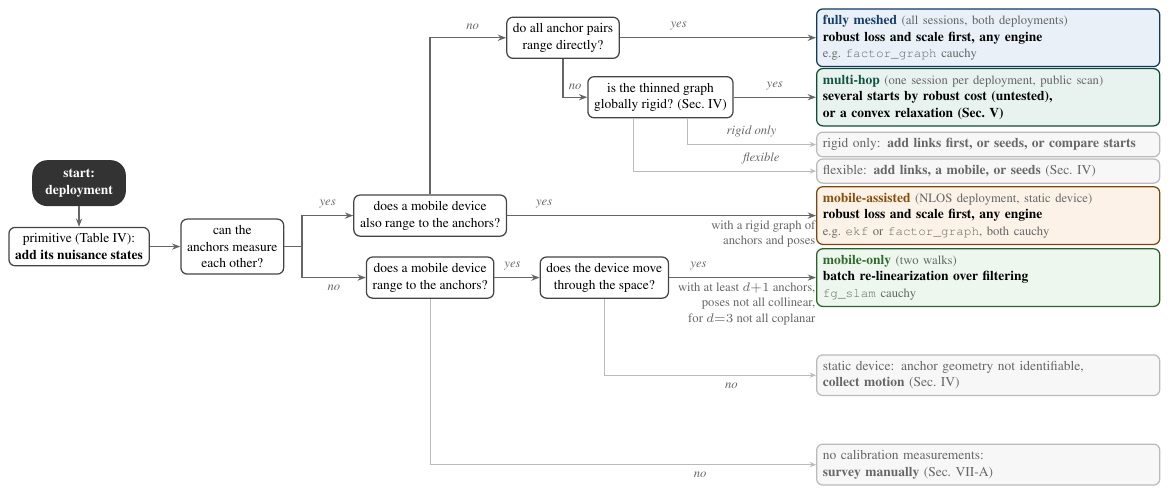}
  \caption{Decision tree from the primitive and edge set of a deployment to a
    topology and an algorithm family. Leaves name their evidence in
    Section~\ref{sec:worked_examples}.}
  \label{fig:decision_tree}
\end{figure*}

Where a method places anchors from the defining nodes of its convention
(Section~\ref{sec:unary}), compare several choices of these nodes at surveyed
check points, or refine the result with a batch estimator, because the choice
sets its error. Rerun with every anchor pair as the first two defining nodes,
the full meshes of \texttt{s03} and \texttt{s21} leave the cauchy factor graph
at one geometry. The \texttt{faithful} closed form places all
anchors for only \WEfcCfFullBthree{} of the \WEfcPairsBthree{} pairs of
\texttt{s03}, at errors from \WEfcCfMinBthree{} to
\SI{\WEfcCfMaxBthree}{\meter}, as the closed forms of the author's prior work
did~\cite{jung_uncertainty_propagation_2025}, and the EKF on \texttt{s21} ends
between \WEfcEkfMinAtwentyone{} and
\SI{\WEfcEkfMaxAtwentyone}{\meter}. The closed form cannot place an anchor
whose ranges to the defining nodes do not close a triangle with them, and its
\texttt{faithful} variant puts every anchor on one side of their line, so a
pair through the interior mirrors the anchors behind it. An error of the range between the
defining nodes moves every anchor, and an error of a range to one of them moves
the anchor placed from it. Seeds fix the absolute frame, as referenced poses or
an alignment to a map~\cite{weber_automatisierte_2021} do, but unlike an
alignment they also act on the shape. With priors of \SI{1}{\centi\meter} on
the first three defining nodes, \texttt{fg\_anchored} (Algorithm~\ref{alg:lm})
brought the unaligned cauchy error close to the blind aligned error on both
full meshes. On the $\kappa=3$ view of \texttt{s03} the seeds raised the error
(Table~\ref{tab:lessons_effect}), since the tight priors and the biased ranges
disagree about a weakly determined shape (Section~\ref{sec:crlb}), and the
$\kappa=3$ view of \texttt{s21} stayed flexible with all three. Seeds cannot
check a calibration, since the priors hold them in place, whereas a single check
point, a surveyed position held out of the estimation, can reveal a large
failure, such as a mirrored frame or a wrong branch, but does not measure the
accuracy.

\section{From Deployment to Method}
\label{sec:decision_tree}

The taxonomy of Section~\ref{sec:taxonomy}, the identifiability conditions of
Section~\ref{sec:identify}, and the engines of Section~\ref{sec:algorithms}
combine with the examples into the decision tree of
Fig.~\ref{fig:decision_tree}. Its questions concern only the deployment
and can be answered before any algorithm choice. Weber's
dissertation~\cite{weber_automatisierte_2021} anticipated it in one dimension,
choosing between MDS and SLAM by the pairwise edge types a deployment can
measure.

The tests of the tree hold for two-way ranges without offset states, and the
counting conditions of Section~\ref{sec:nuisance} extend them to the other
primitives. On a view that is rigid but not globally rigid, links come before
seeds where the ranges are biased (Section~\ref{sec:we_seeds}). On a flexible
view, seeds remove a flex only where they pin nodes that the flex moves
relative to each other. The rule of the multi-hop leaf, several starts
compared by robust cost, is untested here, and the start of lower cost can
reach the worse geometry (Section~\ref{sec:we_thinned}).

For the running example the tree reads as follows. The modules range
peer-to-peer, so the anchors measure each other. Without the tag that stands
still in \texttt{s03}, which the example leaves out, no mobile device ranges to
the anchors, and every anchor pair ranges directly. The path ends at the fully
meshed leaf and its method, the cauchy \texttt{factor\_graph}, which
Section~\ref{sec:we_full_mesh} confirms.
Had the deployment kept only each anchor's three
nearest neighbors, the rigidity test would have returned a rigid but not
globally rigid view and sent it to the terminal that asks for more links
first. The mobile-only view of the walks drops their inter-anchor ranges, and
the tree sends it to the mobile-only leaf and to the cauchy \texttt{fg\_slam}.

\subsection{Calibrate or Survey}
\label{sec:calibrate_or_survey}

A blind calibration and a full survey, which makes calibration non-blind, are
the two ends of the reference-information dimension of
Fig.~\ref{fig:two_axis_space}, with the mobile-referenced and semi-blind
regimes between them. A deployment chooses how much reference
information it needs, and where. Its accuracy target concerns tag positions,
into which every anchor error propagates (Section~\ref{sec:introduction}), so
the anchor errors of Section~\ref{sec:worked_examples} are a proxy. A
survey leaves the range offsets of Section~\ref{sec:nuisance} to
calibrate. The examples test neither how the anchor errors of a calibration
reach position fixes nor offset calibration on a surveyed network. Five
factors bear on the choice.
\begin{itemize}
  \item The accuracy target sets the bar for the anchor errors. The service
    requirements of 5G, for example, set horizontal targets from
    \SI{10}{\meter} down to \SI{0.3}{\meter} for absolute positions and
    \SI{0.2}{\meter} for positions relative to nearby devices, at
    \SI{95}{\percent} confidence~\cite{3gpp_ts_22261_2026}. With a robust
    loss the examples reached anchor RMSEs of decimeters on the full meshes
    (Table~\ref{tab:lessons_effect}). These RMSEs are no 95\,\% levels, so
    they can only be ordered against such targets. A target of meters lies
    above them, and one of decimeters at the level of the best, where a check
    at several surveyed points shows whether a calibration meets it.
  \item The rigidity test of Section~\ref{sec:rigidity} needs only the
    planned edge set, so it can run before any survey. Where it fails, a
    deployment adds links, or surveys or seeds nodes that the flip or flex
    moves, which the test does not name.
  \item An external frame needs $d+1$ seeds, referenced poses, or a map
    alignment (Section~\ref{sec:we_seeds}). A system that needs positions
    only relative to its anchors can stay blind.
  \item The number of anchors and how often they move or change set the cost
    of a survey, which grows with every node and lapses with every relocation
    (Section~\ref{sec:introduction}). A calibration that the infrastructure
    runs itself can be repeated instead, which matters most where surveying
    is impractical, as in emergency response, temporary industrial
    installations, or ad hoc deployments~\cite{raspopoulos_3d_2026}.
  \item Assurance needs a surveyed check point held out of the estimation
    (Section~\ref{sec:we_seeds}) and an uncertainty for the anchors
    (Section~\ref{sec:operational}), above all where integrity or
    certification requirements apply.
\end{itemize}
Calibration serves commissioning and operation alike. In operation, repeating
it can reveal anchors that moved, were replaced, or joined, and so show
whether a survey still holds, a task that Section~\ref{sec:operational} lists
as partly open. These factors usually lead to a hybrid. It surveys $d+1$ seeds
and one check point and calibrates the other anchors with offset states, and
in operation it repeats the calibration. The examples test neither this
combination nor seed priors of other widths.

\section{Open Problems}
\label{sec:discussion}

\subsection{Systematic Bias and Adaptive Noise Scales}
\label{sec:bias_scale}

Range bias that a slightly deformed geometry can absorb is removed neither by
averaging over epochs (Section~\ref{sec:factor_graph}) nor by a robust loss,
so on the LOS deployment it sets the error floor (Section~\ref{sec:we_bias}).
A bias per pair is not identifiable in the blind setting, where it can absorb
any consistent deformation of the geometry~\cite{hamer_self-calibrating_2018}.
Device biases have been estimated before, in a delay stage on a known
geometry~\cite{piavanini_calibration_2022} or jointly with the anchors from
known tag poses~\cite{ortiz_uwb_2023,blueml_bias_2021}. How much bias remains
after offset calibration is open. Because part of the measured bias ends up in
the scale of the estimate (Section~\ref{sec:we_bias}), remedies that carry the
scale, such as one surveyed baseline or a range-dependent bias model
calibrated against a reference~\cite{preter_range_2019}, remain to be tested
on these views.

On the NLOS deployment the session noise scale limits the robust estimators
instead
(Section~\ref{sec:we_full_mesh}). Concomitant scale
estimation~\cite{huber_robust_1964} replaces a fixed scale, and adaptive
robust kernels, whose shape is tuned to the residual
distribution~\cite{chebrolu_adaptive_2021}, replace a fixed kernel. An
adaptive kernel already serves UWB anchor
initialization~\cite{delama_real-time_2025}, and Di~Franco
et~al.~\cite{di_franco_calibration-free_2017} fitted a Gaussian mixture per link
jointly with anchor-free positions from NLOS UWB ranges. Which of these suits
biased ranges without ground truth is open.

\subsection{Testing Identifiability in Sparse and Dynamic Views}
\label{sec:identifiability}

A solver need not find the unique geometry of a generically globally rigid
view~\cite{eren_rigidity_2004,aspnes_theory_2006}. The open problem is a test
for a given instance under robust losses and range bias, since convergence
proves neither uniqueness nor the global minimum, as the flexible and the
random views of Section~\ref{sec:we_thinned} show. Convex relaxations
certify a global optimum where they are tight (Section~\ref{sec:convex}), but
their exactness conditions assume errors without bias. Noise-aware local
tests, such as the robust quadrilaterals of Moore
et~al.~\cite{moore_robust_2004}, avoid placements that noise could flip,
which a generic rigidity test cannot, yet they also leave a persistent bias
out. The information matrix of Section~\ref{sec:crlb} adds the conditioning
that a generic rigidity test lacks. It is evaluated at the surveyed geometry,
which a blind deployment does not have, and ignores per-pair bias, so a
bias-aware test on the measured graph is missing. Since nearest-neighbor
thinning sheds inflated links on the NLOS deployment,
selecting edges that keep a view above its threshold while excluding the
worst links is a planning problem that multi-hop calibration schemes leave
implicit~\cite{herbruggen_multihop_2023,corbalan_self-localization_2023}.
Edge-level fault detection, which Yang et~al.~\cite{yang_beyond_2013} run on
redundantly rigid subgraphs of sensor networks, would add an integrity layer to
the calibration, with the misspecification testing of
Section~\ref{sec:operational} as one candidate.

Mobile-only views replace edge-set conditions with motion conditions. A static
tag leaves the anchor geometry unidentifiable (Section~\ref{sec:motion}), and
the cauchy \texttt{fg\_slam} of Section~\ref{sec:we_mobile} reached its anchor
error with a few thousand epochs of pedestrian motion per walk. For the region
in between, the minimal configurations of ToA self-calibration are
known~\cite{kuang_microphone_2013}, and with known tag poses Blueml
et~al.~\cite{blueml_bias_2021} place triangulation points where the Fisher
information is maximal, while Delama et~al.~\cite{delama_real-time_2025}
start anchor initialization once an online estimate of the position dilution
of precision is low enough. What remains open is a design rule for an unknown
trajectory and a spatially varying bias field that states how much motion a
required anchor accuracy needs.

\subsection{Beyond Two-Dimensional Two-Way Ranging}
\label{sec:measurement_models}

Three measurement models beyond two-way ranging alter the factor residuals
of the examples and add nuisance variables, and each brings a coupling
that two-way ranging avoids. TDoA measurements tie every factor to a reference
node and introduce unknown emission or clock offsets. The acoustic
self-calibration strand has developed the corresponding algebra, including
minimal solvers for TDoA self-calibration~\cite{ask_unifying_2014}. Its
transfer onto radio is direct on the factor graph, as the ToA case on WiFi
round-trip times shows~\cite{batstone_robust_2016}. Where the anchors
synchronize over the air, the clock offset of each anchor contains the time of
flight from the synchronizing node, which has to be known from the node
positions~\cite{tiemann_atlas_2016} or estimated on the same
link~\cite{hamer_self-calibrating_2018}, so bias or NLOS propagation on an
inter-anchor link enters the TDoA
measurements~\cite{zhao_util_2024,friedrich_accurate_2021}. One-way links
between unsynchronized nodes add explicit clock states, which
signals-of-opportunity processing already estimates as a matter of
course~\cite{khalife_opportunistic_2020}. Array-based measurements add
orientation states per anchor, and Bayesian 5G calibration carries both kinds
of state~\cite{lu_bayesian_2023}. None of these requires a new
framework, but the examples cover two-way ranging only, so evaluating
other models across views is open.

Three-dimensional calibration is also open. With coplanar ceiling anchors,
a mobile node's height is weakly observable and ambiguous up to a mirror
about the anchor plane, so this calibration needs height priors or anchors off
that plane.

Both remedies also serve the calibration of the anchors themselves
(Section~\ref{sec:unary}). On measured UWB ranges, a network extended anchor
by anchor accumulated much smaller errors with anchors stacked in pairs at
every second position than with all anchors at one
height~\cite{yasukawa_automatic_2020}. In simulations, measured heights, the
limit of a tight height prior, lowered the error, as did pairs at a known vertical separation once the separation and the number
of surveyed pairs were large enough~\cite{raspopoulos_3d_2026}. On the factor
graph, a pair is one position variable whose second anchor sits at a known
vertical offset. How these remedies compare on the views of
Section~\ref{sec:masking} with ranges measured in three dimensions is open.
The AAU meshed set~\cite{jung_modular_2024,jung_aau_meshed_2024} holds the
edge classes of UWB-ASC, with ground truth for the anchors and motion-capture
truth for an unmanned aerial vehicle, in a strongly three-dimensional anchor
layout, and so offers ranges for that comparison.

\subsection{Learning-Based Methods and Their Evaluation}
\label{sec:learning_eval}

Learned front-ends target the two failure sources the examples expose, NLOS
links and the systematic bias field~\cite{ridolfi_uwb_2021,sun_simultaneous_2025}.
What they can learn depends on what a system logs. A system that logs only
two-way ranges gives one number per directed pair and epoch, as in UWB-ASC,
which records no CIR~\cite{schwarzbach_uwbasc_descriptor_2026}. From ranges
alone, NLOS identification tests the statistics of a link's ranges, such as
their spread against a threshold, and for a static node that spread may not
tell LOS from NLOS~\cite{guvenc_nlos_2008}, so on a static anchor mesh a
learned method has little more than the consistency of a range with the
others to learn from. Where the CIR is logged, classifiers and regressors on
features of the response identify NLOS links and reduce their ranging
error~\cite{marano_nlos_2010}, and the network of Ridolfi et~al.\ predicts the
ranging error from the CIR during anchor
self-calibration~\cite{ridolfi_uwb_2021}. CSI supports channel charting
without any range~\cite{studer_channel_2018}. The edge-masked views of UWB-ASC
therefore test learned methods on ranges alone. Their evaluation is the open
problem. The bias field is a property of a site, so a corrector
trained and evaluated on the same site may memorize local propagation
instead of learning a transferable error model. The two deployments of
UWB-ASC share one hall, so they support no cross-site protocol, and other
sites enter UWB-ASC only through the public recordings its readers
convert.
Trajectory-conditioned
regressors and channel charts are trained per
deployment~\cite{nguyen_uloc_2025,euchner_espargos_2024,yuan_large-scale_2025},
and the gauge analysis of Section~\ref{sec:unary} applies to them
unchanged.

\subsection{Site Knowledge as Reference Information}
\label{sec:site_knowledge}
A site can already hold approximate anchor positions. A building model can
store the positions of installed beacons~\cite{ferreira_beacons_2018}, and a
start close to the true positions, such as a guess by the person who deployed
the nodes, lets range-only estimation reach a small
error~\cite{menegatti_range-only_2009}. The unary prior
of~\eqref{eq:factorization} carries such a position as a noisy seed
(Section~\ref{sec:factor_graph}), and \texttt{fg\_anchored} takes one Gaussian
prior per node at the width that the session gives it, so plan-level seeds need
no new code. Plans age with renovations~\cite{kolakowski_static_2022}, however,
and a Gaussian prior on an anchor mounted away from its planned point pulls the
estimate toward a wrong geometry. Priors with heavy tails, or a test of the
priors against the ranges, remain to be tested. A plan can also fix the frame
of a blind calibration, by matching a recovered trajectory to the map, which
needs a map without axis or point symmetry~\cite{weber_automatisierte_2021},
or through at least three APs whose plan positions are
known~\cite{tong_mapfi_2023}. With such nodes, Abboud
et~al.~\cite{abboud_esnap_2020} assigned the other nodes of a lighting system to
its blueprint room by room, so a matching without known nodes is open. Walls drawn
in a plan can predict which pairs are NLOS, as three-dimensional city models
predict satellite visibility in GNSS shadow
matching~\cite{groves_shadow_2011}, and map-aware localization models cancel
part of the degradation that obstructions cause, more for RSS than for
ToA~\cite{montorsi_map-aware_2014}. Whether such a prediction
can serve calibration as a per-pair prior on bias remains to be shown.

\subsection{Operational and Emerging Aspects}
\label{sec:operational}

A calibration result enters every later position fix, so its uncertainty
belongs in the positioning error budget~\cite{yudanto_device-level_2023}, and
geodetic misspecification testing by detection, identification, and
adaptation~\cite{teunissen_distributional_2018} could exclude faulty edges.
Standards fix how calibrated coordinates reach the positioning function
(Section~\ref{sec:nuisance}), and NR integrity assistance data give devices a
mean and a standard deviation that bound the error of the TRP
locations~\cite{3gpp_ts_38305_2026}. Neither NR nor the location
configuration information of IEEE~802.11~\cite{ieee_802_11_2024} carries the
correlations between anchors that a joint calibration produces. How one
calibration serves several technologies at a shared site, such as UWB anchors
next to APs or small cells, is open as well.

Anchors move, fail, or join during operation, which online initialization of
unknown anchors~\cite{delama_real-time_2025} and incremental
deployment~\cite{buchmayer_factor_2025} address in part. Inter-node ranges
are usually unauthenticated, so a distance-reduction or relay attack during
commissioning can corrupt every later fix, and a manipulation that a slightly
deformed geometry absorbs passes a robust kernel like any consistent bias. The
ranging-integrity features of IEEE 802.15.4z~\cite{ieee_802_15_4z_2020} are
designed to protect the exchange. In 6G networks the infrastructure gains
calibration states, the pose of reconfigurable intelligent
surfaces~\cite{zheng_jrcup_2024} and the phase and time offsets between
distributed antennas under varying prior knowledge of their
positions~\cite{wu_location_2024}, which enter the graph as nuisance states
and priors.

For the released toolkit these problems translate into next steps. The scale
estimators of Section~\ref{sec:bias_scale} can be compared on the released
views, with the scale never selected against the evaluation metric. Running
\texttt{factor\_graph} on the epoch stream that \texttt{ekf} and
\texttt{grid\_filter} read, or these filters on the session medians, would
separate the data interface from the engine (Section~\ref{sec:we_full_mesh}).
A convex initializer and a multi-start of \texttt{factor\_graph} that selects
without truth would test the start beyond the survey
(Section~\ref{sec:we_thinned}).

\section{Conclusion}
\label{sec:conclusion}

Calibrating radio infrastructure from its own measurements is one estimation
problem under many names. This tutorial treats it as a single factor graph
across measurement topologies and kinds of reference information. The lessons
of Section~\ref{sec:worked_examples}, with the effect of each in
Table~\ref{tab:lessons_effect}, and the decision tree of
Section~\ref{sec:decision_tree} lead a deployment to its engine and its
checks. A range bias that varies by pair,
which no blind calibration can identify, heads the open problems of
Section~\ref{sec:discussion}. The code and result tables are public, and the
data are published separately (Section~\ref{sec:records}).

\section*{Data and Code Availability}
\begin{figure}[!htbp]
  \centering
  \includegraphics{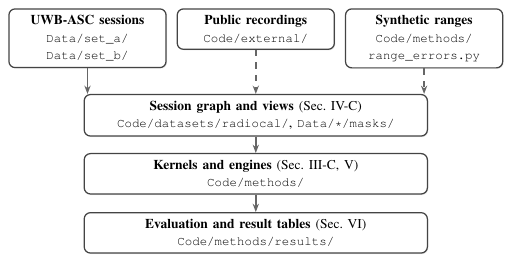}
  \caption{Structure of the release, with the folder of each part and the
    section that explains it. Dashed inputs are not stored under
    \texttt{Data/}.}
  \label{fig:release_structure}
\end{figure}

The examples use UWB-ASC. Its data descriptor documents the deployments and
their ground truth~\cite{schwarzbach_uwbasc_descriptor_2026}, and its Zenodo
record archives the sessions and the raw recordings, with the scripts that
rebuild the measured sessions from them~\cite{schwarzbach_uwbasc_2026}. The
release of this tutorial holds the reference implementations with their
driver, the result tables, the record of the literature search of
Section~\ref{sec:scoping_review}, and a copy of the measured sessions of
UWB-ASC and of its code. The release is published at
\url{https://github.com/PaulSchwarzbach/radiocal}, and the Zenodo deposit
\texttt{10.5281/zenodo.24134343}~\cite{schwarzbach_radiocal_2026} archives it.
Fig.~\ref{fig:release_structure} shows how the release is organized and which
section of the tutorial explains each part.
The code carries the MIT license, and the data, the result tables, and the
literature record carry CC~BY~4.0. The scripts of the release recompute the
result tables from the sessions, those of the synthetic baseline included,
within the tolerances that the README of the release states and apart from
one stored cell on a flexible view that it names. The numbers and figures of
the examples derive from these tables and sessions. Those of the public scan
come from its own files, which stay with the versioned Zenodo record of
TWR-CLOUD-Bern~\cite{xenakis_twr-cloud-bern_2026} under the MIT license. Once
they and the selfloc recordings~\cite{corbalan_self-localization_2023} are
fetched and match the SHA-256 checksums of the fetch list, the scripts under
\texttt{Code/external} of the release convert and evaluate them.
The paper-side scripts, which compute the range residuals, the tables, and the
numbers of the text and draw the figures, write into the manuscript as part of
its build and stay with its sources. The README names for each figure and
table the script and the released files it reads.
One editable install on Python~3.10 to~3.14 runs every script of the release
whose inputs ship with it, and the pinned environment of the result tables
needs Python~3.12 or newer. A smoke test checks an install within seconds, and
a regression gate recomputes the full-mesh cells of two static sessions, the
synthetic baseline, and the reruns of Section~\ref{sec:we_seeds} and compares
them with frozen values. The README gives the commands in order with
their runtimes and states which results agree across environments and within
which tolerance.

\section*{Acknowledgment}
The author used Anthropic's Claude models through the Claude Code
command-line interface~\cite{anthropic_claude_code_2026}. The model versions
were not pinned. The author set and drafted the content of all sections and
the abstract, and the models formulated and edited the text from it. They
helped develop the reference implementations, the evaluation of
Section~\ref{sec:worked_examples}, the conversion of the public scan, and
scripts behind tables and figures. The
author reviewed and revised the text and code and takes full responsibility
for this article.

\bibliographystyle{IEEEtran}
\bibliography{refs}

\end{document}